\documentclass[numberedappendix]{emulateapj}
\usepackage{natbib}
\usepackage{apjfonts}
\usepackage{epsfig}

\newcommand{\LCDM}{\Lambda\mathrm{CDM}}
\newcommand{\Msun}{M_{\sun}}
\newcommand{\Mstar}{M_{\star}}
\newcommand{\Mpc}{\mathrm{Mpc}}
\newcommand{\Gpc}{\mathrm{Gpc}}
\newcommand{\kpc}{\mathrm{kpc}}
\newcommand{\km}{\mathrm{km}}
\newcommand{\s}{\mathrm{s}}

\newcommand{\deltac}{\delta_{c}}
\newcommand{\nuc}{\nu_{c}}
\newcommand{\fsc}{f_{\mathrm{sc}}}
\newcommand{\fec}{f_{\mathrm{ec}}}

\newcommand{\fSMT}{f_{\mathrm{SMT}}}
\newcommand{\aR}{a_{\mathrm{R}}}
\newcommand{\AT}{A_{\mathrm{T}}}
\newcommand{\fT}{f_{\mathrm{T}}}
\newcommand{\gT}{g_{\mathrm{T}}}
\newcommand{\eT}{e_{\mathrm{T}}}
\newcommand{\dT}{d_{\mathrm{T}}}
\newcommand{\hT}{h_{\mathrm{T}}}
\newcommand{\aSMT}{a_{\mathrm{SMT}}}

\newcommand{\vir}{\mathrm{vir}}

\newcommand{\deltasc}{\delta_{\mathrm{sc}}}
\newcommand{\deltaec}{\delta_{\mathrm{ec}}}
\newcommand{\deltaSMT}{\delta_{\mathrm{SMT}}}
\newcommand{\BSMT}{B_{\mathrm{SMT}}}
\newcommand{\Bec}{B_{\mathrm{ec}}}
\newcommand{\Bsc}{B_{\mathrm{sc}}}
\newcommand{\emp}{e_{\mathrm{mp}}}
\newcommand{\pmp}{p_{\mathrm{mp}}}
\newcommand{\dd}{\mathrm{d}}
\newcommand{\dndM}{\dd n/\dd M}
\newcommand{\erfc}{\mathrm{erfc}}

\newcommand{\rhom}{\rho_{m}}

\newcommand{\init}{\mathrm{init}}

\newcommand{\zinit}{z_{\init}}
\newcommand{\tinit}{t_{\init}}
\newcommand{\zc}{z_{\mathrm{c}}}
\newcommand{\rhobarm}{\bar{\rho}_{m}}
\newcommand{\rhobarl}{\bar{\rho}_{\Lambda}}
\newcommand{\vx}{\mathbf{x}}
\newcommand{\What}{\hat{W}}
\newcommand{\Whatk}{\hat{W}_{k}}
\newcommand{\Whatr}{\hat{W}_{\mathrm{r}}}
\newcommand{\Wr}{W_{\mathrm{r}}}
\newcommand{\RW}{R_{\mathrm{W}}}
\newcommand{\Rw}{\RW}
\newcommand{\deltar}{\delta_{\mathrm{R}}}
\newcommand{\deltam}{\delta_{\mathrm{M}}}

\shorttitle{Collapse Barriers and Halo Abundance}
\shortauthors{Robertson et al.}

\begin{document}

\title{Collapse Barriers and Halo Abundance: Testing the Excursion Set Ansatz}
\author{Brant E. Robertson\altaffilmark{1,2,5}, Andrey V. Kravtsov\altaffilmark{1,2},
Jeremy Tinker\altaffilmark{1,4}, and Andrew R. Zentner\altaffilmark{3}}

\altaffiltext{1}{Kavli Institute for Cosmological Physics, and Department of Astronomy and
Astrophysics, University of Chicago, 933 East 56th Street, Chicago, IL 60637, USA}
\altaffiltext{2}{Enrico Fermi Institute, 5640 South Ellis Avenue, Chicago, IL 60637, USA}
\altaffiltext{3}{Department of Physics and Astronomy, University of Pittsburgh, Pittsburgh, PA 15260, USA}
\altaffiltext{4}{Current Address: Berkeley Center for Cosmological Physics, University of California, Berkeley}
\altaffiltext{5}{Spitzer Fellow}

\begin{abstract}
Our heuristic understanding of the abundance of dark matter halos centers around the concept
of a density threshold, or ``barrier'', for gravitational collapse.  If one adopts the
{\it ansatz} that regions of the linearly evolved density field smoothed on mass scale $M$ with an overdensity that exceeds the 
barrier will undergo gravitational collapse into halos of mass $M$, the corresponding abundance of such
halos can be estimated simply as a fraction of the mass density satisfying the collapse criterion divided by 
the mass $M$. The key ingredient of this {\it ansatz} is therefore the functional form of the
collapse barrier as a function of mass $M$ or, equivalently, of the variance $\sigma^2(M)$. 
Several such barriers based on the spherical, Zel'dovich, and ellipsoidal collapse models have been 
extensively discussed. Using large-scale cosmological simulations, we show that the
relation between the linear overdensity 
and the mass variance for regions that collapse to form halos by the present epoch
resembles expectations from dynamical models of ellipsoidal collapse.  
However, we also show that using such a collapse barrier with
the excursion set {\it ansatz} predicts a halo mass function inconsistent with that measured directly 
in cosmological simulations. 
This inconsistency demonstrates a failure of the excursion set {\it ansatz} as a physical
model for halo collapse.
We discuss implications of our results for understanding the collapse epoch for
halos as a function of mass, and avenues for improving consistency between
analytical models for the collapse epoch and the results of cosmological simulations.
\end{abstract}

\section{Introduction}
\label{section:introduction}

A central concept in the modern theory of galaxy formation 
is the connection between characteristics of the 
linear density field and the abundance and properties
of virialized dark matter halos 
in the contemporary universe.  
The power spectrum of density perturbations seeded by inflation
\citep[e.g.,][]{guth1982a,bardeen1983a,starobinsky1983a}
and the cosmological transfer function
\citep[e.g.,][]{peebles1982a,bardeen1986a,eisenstein1998a}
determine the character of the subsequent
nonlinear growth of structure through gravitational
clustering \citep{white1978a}. Growing perturbations in the initial 
density field serve as the sites of galaxy formation
\citep[e.g.,][]{peebles1965a,sachs1967a,white1978a}.  In the context of a
cold dark matter cosmology, these processes give
rise to the characteristic mass scale of observed
galaxies \citep{rees1977a,blumenthal1984a}.  Cosmological observations
have both motivated and verified this picture, most
recently with measurements of galaxy clustering
\citep[e.g.,][]{percival2007a}, the linear power spectrum of cosmological
structures \citep[e.g,][]{mcdonald2006a}, and
high-precision measurements of the cosmological
microwave background radiation \citep[e.g.,][]{dunkley2008a}.

Methods for calculating the abundances of nonlinear, 
collapsed structures have been developed
to link the growth of density perturbations
with the observed number densities of galaxy- and cluster-scale
objects.  Dynamical models for the collapse of individual 
dense patches into virialized structures, such as the spherical collapse 
\citep[][see Appendix \ref{section:spherical_collapse}]{gunn1972a} and ellipsoidal collapse 
\citep[][see Appendix \ref{section:ellipsoidal_collapse}]{eisenstein1995a,bond1996a} models,
provide physically-motivated methods for estimating
the necessary, linearly-extrapolated overdensity (the
``collapse barrier'') for a region to break from the cosmic expansion, 
condense, and form a high-density, virialized structure (i.e., a
dark matter halo).  When combined with the statistics of the
initial density field, the collapse barrier can thereby be utilized to
estimate the abundance of dark matter halos as a function of
mass and redshift.  The purpose of this paper is to 
re-examine the connection between the collapse barrier and halo
abundance, and test the common assumptions and methodologies used to
calculate the mass function of dark matter halos from a dynamical
model for their collapse~\citep[e.g., the ``excursion set'' formalism,][]{bond1991a}.

\cite{press1974a} first used the spherical collapse model to 
calculate the abundance of galaxies.  
They assumed the probability distribution function $\dd P(\deltar)/\dd \deltar$ (PDF)
of the smoothed overdensity field $\deltar = (\rhom-\rhobarm)/\rhobarm$, where the 
mean matter density is $\rhobarm$ and the density is averaged over 
a region of typical size $R$ containing mass $M \propto \rhobarm R^3$,
was a Gaussian with a scale-dependent variance $\sigma^{2}(M)$.  
They 
integrated this Gaussian PDF above the typical collapse overdensity $\deltac$ 
(i.e., $P(\deltam > \deltac) \propto \erfc[-\deltac/\sqrt{2}\sigma(M)]$)
and differentiated with respect to mass $M$ 
(i.e., $\dd P/\dd M \propto \exp[-\deltac^{2}/2\sigma^{2}(M)]\times\dd \sigma^{-1}/\dd M$)
to arrive at the fraction of all mass contained in objects of mass $M$.  
The \citet{press1974a} calculation accounts for only half of the total universal mass density 
in bound objects because it does not address underdense regions contained within still 
larger regions for which the threshold $\deltam > \deltac$ is satisfied.  To remedy this shortcoming, 
\citet{press1974a} multiplied their final answer by a factor of two to account for all mass 
with little justification.

\citet{bond1991a} studied the properties of sets of regions above the 
threshold, the ``excursion sets'' of the density field.  
Using the excursion set formalism, \cite{bond1991a} 
demonstrated that by filtering the initial overdensity  
field on a variety of mass scales, the results of
\cite{press1974a} could be derived in a manner  
that accounts for patches of low density  
embedded in large, high-density regions collapsing on 
larger scales (the ``cloud-in-cloud'' problem).  
Moreover, \cite{bond1991a} and \cite{lacey1993a}
showed that the excursion set formalism provided a means to compute other 
halo properties such as their mass acquisition histories.  
The excursion set theory of halo abundance 
was later extended by \cite{mo1996a} to describe their
spatial clustering through the bias parameter $b^{2}\equiv\xi_{hh}/\xi_{m}$
relating the halo ($\xi_{hh}$) and mass ($\xi_{m}$) 
correlation functions \citep[a computation that recovers the 
``peak-background split'' result developed in][]{kaiser1984a,efstathiou1988a,cole_kaiser1989}.  
The details of the excursion set theory are collected in the recent review by \citet{zentner2007a}.

Numerical simulations of cosmological structure formation, 
which were developed concurrently with
the analytical collapse calculations,
demonstrated that nonlinear gravitational collapse 
produces halo mass functions that are inconsistent
with the predictions of \cite{press1974a}.
Evidence for this disagreement, as well as corresponding
discord in the spatial clustering of halos, developed over twenty years 
\citep[e.g.,][]{efstathiou1988a,white1993a,lacey1994a,eke1996a,gross1998a,tormen1998a,jing1998a,lee1999a,jing1999a,porciani1999a,governato1999a}.
The inability of the simple, spherical collapse model set within 
the excursion set formalism to describe the results of cosmological
simulations and the need for robust predictions
of the abundance of halos for comparisons
with observations led to accurate formulae 
for mass functions determined by
numerical fits to the results of N-body simulations
\citep[e.g.,][for more recent results see \citealt{tinker2008a}]{sheth1999a,jenkins2001a,warren2006a}.
These fitting formulae were not the
results of specific dynamical models for the collapse
of dark matter halos.  Rather, they were developed 
through a practical approach of trying to reproduce accurately 
the results of cosmological simulations.

Additional dynamical models for the growth of structure were developed
in an attempt to better reproduce simulated mass functions while retaining 
a comparably simple, physically-motivated framework.
\cite{lee1998a} presented results for the
halo mass function motivated by the \cite{zeldovich1970a}
pancake collapse model, which they extended 
to account for the collapse of halos along each of their principal axes.
\citet[][SMT01]{sheth2001a} argued that the
functional form of the \citet[][ST99]{sheth1999a} mass function
can be motivated by the ellipsoidal collapse model of \cite{bond1996a}.
For a given ellipticity $e$ and prolaticity $p$
of the shear field about an overdensity $\delta$,
the \cite{bond1996a} model provides a method for estimating the 
linearly-extrapolated overdensity at collapse
(the ``ellipsoidal collapse barrier'').
SMT01 built on a calculation by \cite{doroshkevich1970a}
to determine a probability distribution of shear ellipticities
and prolaticities, which was then used to find the most-probable 
values for $e$ and $p$ as a function of $\sigma(M)$.
Combining these results, SMT01
found an effective ellipsoidal
collapse barrier as a function of the ``peak height'', 
$\nu_{c}=\delta_{c}/\sigma(M)$, alone.   With this new, 
mass-dependent collapse barrier  
SMT01 computed a halo mass function using the excursion set prescription of  
\citet{bond1991a}
and showed that the predicted functional form was close to that measured in cosmological
simulations by ST99. The ST99 mass function has
therefore become associated with the ellipsoidal collapse
model.

Note, however, that the ellipsoidal collapse model predicts that the 
collapse barrier converges to the spherical collapse barrier $\deltac=1.69$ (for $\Omega_{m}=1$), 
in the high-mass limit, because the rarest peaks have a preferentially spherical shape.
SMT01, on the other hand, found that the collapse barrier had
to be lowered in the high-mass limit to $\sqrt{a_{\mathrm{ST}}} \deltac \sim 0.84 \deltac$, 
in order to reproduce the mass 
function measured in cosmological simulations \citep[see also][]{sheth2002a}. Although
they argued that the lower value can be motivated by the mass definition of the Friends-of-Friends
algorithm which they used to identify halos in simulations, this rescaling is not well 
justified.  Indeed, it disagrees with the fact that the collapse of the highest mass
halos should be described well by the spherical collapse model. We argue below 
the lowering of the collapse barrier is instead required by the internal inconsistencies
of the excursion set {\it ansatz} and 
explicitly demonstrate that such inconsistencies persist for a wide variety of
halo mass definitions, including the Friends-of-Friends and spherical overdensity criteria 
(see Appendix \ref{section:mass_definitions} for a detailed discussion).

In the context of the excursion set formalism, the adoption
of a shape for the collapse barrier 
(along with choice of prescription with which to smooth the density field)
effectively determines the mass function.  In this paper we test 
the excursion set {\it ansatz} by measuring the effective collapse barrier for halos
formed in cosmological simulations and comparing excursion set predictions 
for the halo abundance given such a barrier to the halo mass functions measured
in the simulations.

The organization of this paper is as follows.
In \S \ref{section:theory},
we review the theoretical background of the excursion set formalism,
common dynamical models, and collapse barriers from the literature.
In \S \ref{section:simulations},
we measure the linear overdensity of collapsed regions 
in cosmological simulations to study the consistency
between the simulated collapse of halos and the predictions
of dynamical models.
We then compare the abundance of dark matter halos predicted
from the excursion set formalism and dynamical models with
the simulated halo mass function in \S \ref{section:fcds}.
We discuss our results in \S \ref{section:discussion} and
summarize our results and conclusions in \S \ref{section:summary}.
The paper contains four appendices: a review of the
spherical collapse model (Appendix \ref{section:spherical_collapse}),
a review of the \cite{bond1996a} ellipsoidal collapse
model (Appendix \ref{section:ellipsoidal_collapse}), 
a summary of the \cite{zhang2006a} analytical method
for calculating the excursion set mass function
(Appendix \ref{section:zh06}), and a study of the
connection between the excursion set {\it ansatz }
and the halo mass definition (Appendix \ref{section:mass_definitions}).
Unless noted otherwise, throughout the paper we
assume a flat $\Lambda$-Cold Dark Matter ($\LCDM$) cosmology.  Cosmological parameters in our calculations and
simulations are close to the values suggested by observations:
$\Omega_{m}\approx 0.3$, $\Omega_{\Lambda}\approx0.7$, and $H_{0}\approx70~\km~\s^{-1}~\Mpc^{-1}$.

\section{Theoretical Background}
\label{section:theory}

Consider a Gaussian random density field 
$\rho(\vx)$ as a function of spatial 
location $\vx$
with mean matter density $\rhobarm$.
At every location $\vx$ we can define the
overdensity $\delta(\vx) \equiv [\rho(\vx)-\rhobarm]/\rhobarm$
of the field and the overdensity
smoothed on a comoving length scale $\Rw$ as
\begin{equation}
\label{equation:smoothed_overdensity}
\delta_{\RW}(\vx) = \int \dd^{3} x' \delta(\vx') W(|\vx-\vx'|,\RW)
\end{equation}
where $W(\vx,\RW)$ is a spherically-symmetric smoothing window with a 
characteristic radius $\RW$ centered about location $\vx$.  Here, we follow the notation in the
recent review by \cite{zentner2007a} but use the symbol 
$\delta_{\RW}$ to represent the overdensity field smoothed on a scale $\RW$ and
reserve $\delta$ to represent the unsmoothed overdensity field.  
The smoothed overdensity field $\delta_{\RW} (\vx)$ is also a 
Gaussian random field with a variance given by
\begin{equation}
\label{equation:sigma}
S(M) \equiv \sigma^2(M) \equiv \left<\delta^{2}_{\RW}(\vx)\right>= \int \Delta^{2}(k) |\What(k,\Rw)|^{2} \dd \ln k,
\end{equation}
where $\Delta^{2}(k) = k^{3}P(k)/2\pi^{2}$ is the dimensionless power 
spectrum of density fluctuations with a wavenumber $k$ and 
$\What(k,\Rw)$ is the Fourier transform of the
real-space smoothing window $W(\vx,\Rw)$.  
The details of both the averaging procedure and the mass-radius relation 
are fixed by the choice of filtering function used to smooth the density field.  
In CDM models the variance monotonically decreases with increasing length or mass scale.  
Consequently, once the filter function is specified 
this variance can also be used to label the size of the 
smoothing region, and we can write $\delta_{\RW}$ as $\delta_{M}$ or 
$\delta_S$ where $M$ is the mass contained within the window of length scale $\RW$ 
and $S$ is the variance $\sigma^2(M)$.
In what follows, we adopt the
common practice of labeling the size of the smoothing region 
by either the length scale $\RW$, the corresponding 
mass scale $M$, or the variance $S$,
and use the labels interchangeably. 

If we consider a fixed location $\vx$ and monitor the behavior
of the smoothed overdensity $\delta_S(\vx)$ as we decrease the mass
smoothing scale $M$ from some very large value 
(and hence increase the variance $S$ from some value $\ll \deltac^2$),
$\delta_S(\vx)$ will execute a (not necessarily Markovian)
random walk where the smoothed overdensity $\delta_{S}(\vx)$ will 
satisfy the Langevin equation \citep[e.g.,][see also \citealt{chandrasekhar1943a}]{bond1991a}
\begin{equation}
\label{equation:langevin}
\frac{\partial\delta_S(\vx)}{\partial\ln k} = Q(\ln k)\What(k,\RW),
\end{equation}
\noindent
where $Q(\ln k)$ is a Gaussian random variable with zero mean and variance
\begin{equation}
\label{equation:langevin_variance}
\left<Q^{2}(\ln k)\right> = \frac{\dd S}{\dd \ln k} = \Delta^{2}(k).
\end{equation}
\noindent
The variation represents an ensemble of local realizations of the density field.

For a sharp $k$-space tophat filter of the form
\begin{equation}
\label{equation:filter_k_space_tophat}
\Whatk(k,\Rw) = \Theta(1 - k\Rw),
\end{equation}
\begin{equation}
\label{equation:heaviside_step_function}
\Theta(x) \equiv \left\{\begin{array}{r@{\quad:\quad}l} 0&x<0 \\ \frac{1}{2}&x=0 \\ 1&x>0 \end{array}\right.,
\end{equation}
\noindent
the ``trajectory'' of overdensity as a function of variance $\delta_S(\vx)$ 
recovered by integrating Equation \ref{equation:langevin} will be a 
Markovian random walk because each Fourier coefficient of the Fourier-transformed 
density field is an independent random variable \citep[e.g.,][]{lacey1993a}.  
More generally, the trajectory may vary more smoothly 
as the variance is increased if the Fourier transform of the 
real-space window function has broad side lobes
in $k$-space \citep[e.g.,][]{zentner2007a}.  
For instance, the real-space tophat filter, given by
\begin{equation}
\label{equation:filter_real_space_tophat}
\Wr(R,\Rw) = \left(\frac{4\pi}{3}R^{3}\right)^{-1}\Theta(1 - R/\Rw),
\end{equation}
with Fourier transform
\begin{equation}
\label{equation:filter_real_space_tophat_transform}
\Whatr(k,\Rw) = \frac{3\left(\sin k\Rw - k\Rw\cos k\Rw\right)}{\left(k\Rw\right)^{3}},
\end{equation}
\noindent
has extended side lobes in $k$-space that correlate the smoothed 
overdensities over a considerable range of smoothing scales.  Hence,
the integral of Equation \ref{equation:langevin} will vary more
smoothly with the variance $S$ if a real-space tophat is used rather
than a $k$-space tophat.

\subsection{Collapse Barriers and Halo Formation}
\label{section:theory:barriers}

The excursion set theory of halo abundance, clustering, and 
formation is based on an {\it ansatz} that the locations and sizes 
of virialized dark matter halos can be related to the properties of peaks in 
the initial density field at some very high initial redshift $\zinit \gg 1$, 
when the density field is in the linear regime ($\delta_{\rm M}\ll 1$).  
Almost universally, the specific form of this {\it ansatz} is that a region will 
collapse and form a dark matter halo if its smoothed overdensity, evolved forward 
in time from $\zinit$ according to linear perturbation theory, exceeds some 
threshold value.  Consider an object that collapses at some redshift $\zc < \zinit$.  
The linearly-extrapolated overdensity is 
\begin{equation}
\label{equation:collapse_redshift}
\delta_{\RW}(\vx,\zc)=\delta_{\RW}(\vx,\zinit)D(\zc)/D(\zinit), 
\end{equation}
where $D(z)$ is the linear growth function, given by 
\begin{equation}
\label{equation:growth_function}
D(z) = D_{0}H(z)\int_{z}^{\infty} \frac{\left(1+z'\right)\dd z'}{H^{3}(z')}.
\end{equation}
The Hubble parameter
\begin{equation}
\label{equation:hubble_parameter}
H(z) = H_{0}\left[\Omega_{m}(1+z)^{3} + (1-\Omega_m -\Omega_{\Lambda})(1+z)^{2} + \Omega_{\Lambda}\right]^{1/2}
\end{equation}
\noindent
describes the rate of change of the universal scale factor
as $H\equiv\dot{a}/a$.  The collapse condition is then simply
\begin{equation}
\label{equation:collapse_barrier}
\delta_{\RW}(\vx,\zinit)D(\zc)/D(\zinit) \ge B,
\end{equation}
where $B$ is referred to as the ``collapse barrier.''  The excursion set 
{\it ansatz} is specifically that the largest smoothing scale $\RW \propto (M/\rhobarm)^{1/3}$ 
at any point $\vx$ for which Equation \ref{equation:collapse_barrier} is satisfied 
will collapse and form a halo of mass $M$ at redshift $\zc$.

The value of the collapse barrier $B$ is usually determined by a dynamical model for the 
collapse of overdense patches in a background cosmological environment.  In general, 
the collapse barrier need not be a single number and $B$ can be a complicated function of 
the properties of the local linear density field (including its spatial derivatives), 
the smoothing window, and the smoothing scale.
In the following sections, 
we study the spherical collapse 
and ellipsoidal collapse models for $B$.

Having specified the collapse condition and the form of the collapse barrier, 
the number of collapsed objects at a given mass $M$ or variance $S$
will be determined by the probability distribution $f(S)\dd S$ of 
variances where random realizations of trajectories 
$\delta_S(\vx)$, computed according to Equation \ref{equation:langevin}, 
first cross the barrier $B$.  This first-crossing distribution can 
be determined using a Monte Carlo procedure.  By integrating Equation \ref{equation:langevin} 
for many locations in the density field [corresponding to many
realizations of $Q(\ln k)$], the first-crossing distribution
may be approximated by a histogram of the
barrier crossings as a function of $S$ for the ensemble of trajectories.
The first-crossing distribution is often written 
as a function of the peak height 
\begin{equation}
\label{equation:nuc}
\nuc=\delta_{c}/\sigma(M),
\end{equation}  
where $\delta_{c}=1.686$ is the
linear overdensity for spherical collapse in an $\Omega_{m}=1$
cosmology (see \S \ref{section:spherical_collapse} below for details).  
We will frequently change variables from $S$ to $\nuc$ so that we can
discuss and plot the first-crossing distribution in terms of $f(\nuc)$.

The comoving abundance of halos $(\dndM)\times \Delta M$ in the mass
range $\Delta M$ about mass $M$ (the ``mass function'') is related to
the first-crossing distribution $f(\nuc)$ by
\begin{equation}
\label{equation:mass_function}
\frac{\dd n}{\dd M}\Delta M = \frac{\rhobarm}{M} f(\nuc) \left|\frac{\dd \nuc}{\dd M}\right|  \Delta M.
\end{equation}
\noindent
This correspondence between the mass function
and first-crossing distribution can be understood as a
variable change from a distribution in peak height $\nuc$ to a 
distribution in mass $M$, with a normalization that accounts for
the partitioning of mass elements into halos of mass $M$ (i.e., $\rhobarm/M$).
The shape of the first-crossing distribution $f(\nuc)$ 
is expected to be relatively independent of the cosmological model at low redshift 
because it depends primarily on the primordial power spectrum, the collapse barrier $B$, 
and the smoothing window (e.g., ST99).
The halo abundance $\dd n/\dd M$ additionally depends on how the relation between
the trajectory mass smoothing scale $M$ and the actual halo mass is defined
and the shape of the power spectrum $P(k)$ (through $\dd \sigma^{2}/\dd M$). 
When necessary, we will associate halos with roughly virialized regions 
of size $R_{200}$ with a mean physical overdensity $\Delta= 200$
\citep[to ease comparison with the mass definition used by][which is defined 
relative to the background density, and not the critical density]{tinker2008a}
and virial mass $M_{200} = 4\pi\Delta\rhobarm R_{200}^{3}/3$.  However, we explore 
variations on this definition for completeness (see Appendix \ref{section:mass_definitions}).

\begin{figure}
\figurenum{1}
\epsscale{1.0}
\plotone{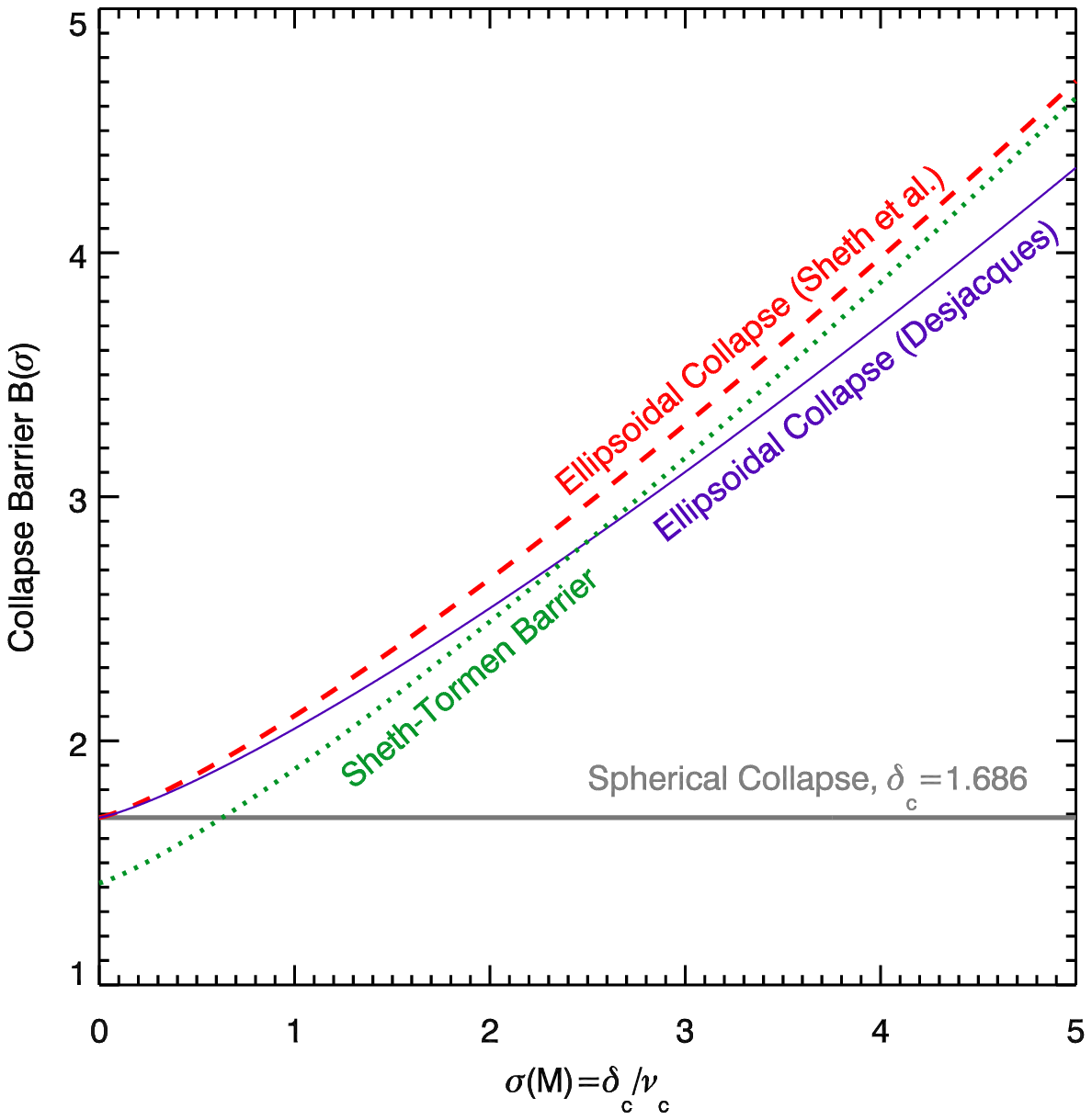}
\caption{\label{fig:barriers}
Some common collapse barriers discussed in the text: the spherical collapse
barrier \citep[][$\deltac\approx1.686$ for $\Omega_{m}=1$, solid gray line]{gunn1972a},
the ellipsoidal collapse
barrier \citep[][dashed red line]{sheth2001a},
an alternative fit to the ellipsoidal collapse
barrier \citep[][solid purple line]{desjacques2008a},
and
the modified ellipsoidal collapse
barrier associated with the \cite{sheth1999a} mass function \citep[][dotted green line]{sheth2001a}.
Note that while the ellipsoidal collapse barriers, as formulated, converge to spherical collapse
barrier at large masses [small $\sigma(M)$], 
the modified barrier of \cite{sheth2001a} converges to a lower value of $\sqrt{a_{\rm ST}}\deltac\approx0.84\deltac$.
\\\\\\
}
\end{figure}

\subsection{Spherical Collapse Barrier}
\label{section:theory:spherical_collapse_barrier}

\cite{gunn1972a} modeled the dynamical evolution
of an overdense spherical region in a background  
cosmology (their calculation is detailed in Appendix \ref{section:spherical_collapse}).  
The collapse barrier 
is computed by evolving the density in the spherical region 
according to the linear theory until the time of collapse.  
The final step in calculating the excursion set halo abundance 
from the collapse model 
is to assume that this equivalent linear overdensity
can be used to identify collapsed regions in the initial density field 
without needing to consider the field's nonlinear evolution.
The model predicts that a region with initial physical overdensity
$\delta_S(\zinit)$ will collapse when the linearly-extrapolated overdensity 
exceeds   
\begin{equation}
\label{equation:spherical_collapse_barrier}
\delta_S(z) = \delta_S(\zinit) D(z)/D(\zinit) > \Bsc \equiv \delta_{c}\approx1.686.
\end{equation}  
We refer to the barrier $\deltac$ as the ``spherical collapse barrier.''  
\cite{eke1996a} provide analytical solutions for the value of $\delta_{c}$
in cosmologies with $\Omega_{m}+\Omega_{\Lambda}=1$.
The spherical collapse barrier is independent of mass scale 
and initial overdensity.

\cite{bond1991a} used the excursion set formalism to
calculate the first-crossing distribution associated
with a barrier that is constant as a function of mass scale, such as the 
spherical collapse barrier.  For a sharp-$k$
window function [Equation~(\ref{equation:filter_k_space_tophat})],
the spherical collapse barrier first-crossing distribution is 
\begin{equation}
\label{equation:spherical_collapse_fcd}
\nuc \fsc(\nuc) = 2 \left(\frac{\nuc^{2}}{2\pi}\right)^{1/2}\exp\left(-\frac{\nuc^{2}}{2}\right).
\end{equation}
\noindent
The function $\fsc$ is normalized such that $\int \fsc(\nuc)\dd\nuc=1$, implying 
that all mass in the universe is incorporated into collapsed objects.  In this context, 
the normalization arises because the variance $S=\sigma^2(M)$ increases monotonically 
toward infinity as $M$ tends toward zero, while the barrier height $\deltac$, remains fixed.  
As a result, any random walk will cross the barrier $\deltac$ at some scale.

\subsection{Ellipsoidal Collapse Barrier}
\label{section:theory:ellipsoidal_collapse_barrier}

The spherical collapse model is likely too simplistic because peaks
in the linear density field are, in general, locally triaxial
\citep[see, e.g., ][]{doroshkevich1970a,bardeen1986a}.  A number of
dynamical collapse models designed to account for deviations from
spherical symmetry have been explored
\citep[e.g.,][]{zeldovich1970a,nariai1972a,hoffman1986a,bertschinger1994a,eisenstein1995a,bond1996a,audit1997a,del_popolo2001a,shen2006a}.
\cite{bond1996a} studied an ellipsoidal collapse model
that approximates peaks in the linear density field as 
ellipsoids and accounts for the effects of tides on the evolution of overdense patches
(see Appendix \ref{section:ellipsoidal_collapse} for a more detailed review of the model).
Compared with the spherical case,
the key feature of the ellipsoidal collapse models 
is that for a fixed overdensity the collapse epoch will
depend on the local ellipticity and prolaticity of the shear (or possibly density) field.  
The net effect is that less spherical peaks have to overcome additional 
tidal stretching and require a higher overdensity to collapse.  

SMT01 used the \cite{bond1996a} ellipsoidal collapse model to
derive the dependence of collapse overdensity on the ellipticity and
prolaticity of the shear field. They found that the barrier shape could be 
approximated by the solution of the implicit equation
\begin{equation}
\label{equation:deltaec_ep}
\frac{\deltaec}{\deltac} = 1 + \beta\left[5(e^{2}\pm p^{2})\left(\frac{\deltaec^{2}}{\deltac^{2}}\right)\right]^{\gamma},
\end{equation}
\noindent
where $\beta$ and $\gamma$ are numerical parameters 
that must be fit to the results of the dynamical
model and the squares of the ellipticity, $e$, and prolaticity, $p$,
are summed (differenced) if $p<0$ ($p>0$).

Determining the shape of the ellipsoidal collapse
barrier requires a further model 
for how the typical ellipticity or prolaticity
scales with galaxy mass because the barrier shape depends 
explicitly on both $e$ and $p$.  
SMT01 use the results of \cite{doroshkevich1970a}
to arrive at a probability distribution function for
$e$ and $p$, from which they find the most probable 
prolaticity and ellipticity are $\pmp=0$
and $\emp=\sigma/\delta\sqrt{5}$, respectively.
They then set $\delta=\deltaec$ and
substituted $\emp$ and $\pmp$
into Equation~(\ref{equation:deltaec_ep}) to find
\begin{equation}
\label{equation:ellipsoidal_collapse_barrier}
\Bec \equiv \deltaec = \deltac\left[ 1 + \beta\left(\frac{\sigma^{2}(M)}{\deltac^{2}}\right)^{\gamma}\right].
\end{equation}
SMT01 found the parameter values $\beta \approx 0.47$ and $\gamma \approx 0.615$ 
\citep[more recently, values of $\beta\approx0.412$ and $\gamma\approx0.618$ were found by][]{desjacques2008a}.
We refer to the threshold in Equation~(\ref{equation:ellipsoidal_collapse_barrier}) 
as the ``ellipsoidal collapse barrier.''
For reference, Figure \ref{fig:barriers} compares the ellipsoidal
collapse barrier with the constant spherical collapse barrier.

SMT01 also suggested an analytical form to approximate
the first-crossing distribution resulting from the barrier $\deltaec$, which 
they calculated numerically using Monte Carlo realizations of the Langevin 
equation [Equation~(\ref{equation:langevin})].  They prescribed the formula
\begin{equation}
\label{equation:ellipsoidal_collapse_fcd}
\nuc \fec(\nuc) = 2 A \left(1 + \nuc^{-2q}\right)\left(\frac{\nuc^{2}}{2\pi}\right)^{1/2}\exp\left(-\frac{\nuc^{2}}{2}\right),
\end{equation}
\noindent
where $q=0.3$, and the constant $A=0.3222$ is determined by requiring
$\int \fec(\nuc)\dd\nuc=1$.

\subsection{The Sheth et al.~Barrier and Mass Function}
\label{section:theory:sheth_tormen_barrier}

To improve the agreement between the excursion set mass function determined by
using Equation (\ref{equation:ellipsoidal_collapse_barrier}) and the 
abundance of halos in the GIF simulations \citep{kauffmann1999a} measured by ST99,
SMT01 introduced another parameter, 
$\aSMT\approx 0.707$, to modify the  
ellipsoidal collapse barrier as
\begin{equation}
\label{equation:gif_simulation_barrier}
\BSMT\equiv\deltaSMT = \sqrt{\aSMT}\deltac\left[ 1 + \beta\left(\frac{\sigma^{2}(M)}{\aSMT\deltac^{2}}\right)^{\gamma}\right],
\end{equation}
\noindent
and changed the values of the other parameters to $\beta=0.5$ and
$\gamma=0.6$.  We refer to Equation~(\ref{equation:gif_simulation_barrier}) 
as the ``Sheth et al.~barrier,'' to contrast it with 
the ellipsoidal collapse barrier because the changes 
are not based on the dynamical collapse model.  

The modified 
barrier is lower than the $\Bec$, with the difference increasing
with increasing mass. In particular, instead of converging to the spherical collapse
barrier of $\deltaec\to\deltac$ at the largest masses, as
expected from the trend towards sphericity for the rarest peaks,
the modified barrier converges to
$\deltaSMT\approx\sqrt{\aSMT}\deltac\approx0.84\deltac$.  The value
$\aSMT=0.707$ was justified by 
SMT01 
as accounting for the
particular choice of the halo mass definition in the GIF simulations
(identified by the Friends-of-Friends algorithm with a linking
length $b=0.2$; see the discussion in \S 4.1 of SMT01 and 
Appendix \ref{section:mass_definitions}).
For
reference, the Sheth et al.~barrier is plotted in Figure~\ref{fig:barriers} alongside 
the spherical collapse and ellipsoidal collapse barriers.

SMT01 utilized Equation~(\ref{equation:gif_simulation_barrier}) 
as the collapse barrier to calculate an
excursion set mass function using
Monte Carlo methods \citep[see also][]{sheth2002a}, and found the
resulting first-crossing distribution to be well-approximated by the
analytical formula
\begin{eqnarray}
\label{equation:sheth_tormen_fcd}
\nuc \fSMT(\nuc) &=& 2 A \left[1 + \left(\sqrt{\aSMT}\nuc\right)^{-2q}\right]\left(\frac{\aSMT\nuc^{2}}{2\pi}\right)^{1/2}\nonumber\\
&\times&\exp\left(-\frac{\aSMT\nuc^{2}}{2}\right),
\end{eqnarray}
\noindent
with $\aSMT=0.707$, $q=0.3$, $A=0.3222$.
This formula was introduced by
ST99 as a fit
to the GIF simulations, and
the corresponding halo mass function [calculated
from Equation~(\ref{equation:mass_function})]
is often referred to as the ``Sheth-Tormen'' mass function.
The normalization of the first-crossing
distribution for the Sheth-Tormen
mass function is the same as the
spherical and ellipsoidal collapse first-crossing
distributions, with $\int \fSMT(\nuc)\dd \nuc=1$.

\section{Testing the collapse barrier with cosmological simulations}
\label{section:simulations}

As we have noted above, the two main ingredients of the 
excursion set formalism are the collapse barrier used to 
decide which mass elements form halos and 
the sampling of the possible collapse histories
via the random walks of the smoothed overdensity field.  
Each of the elements of the 
excursion set approach can be tested against cosmological numerical 
simulations of structure formation.  In particular, the correct collapse 
barrier, if indeed one can be defined, may be verified or falsified 
using numerical simulations that include all of the complications that 
the excursion set approach aims to circumvent.
In the previous section, we described the collapse barriers
motivated by the spherical and ellipsoidal collapse models, as well
as the modified barrier of SMT01.
In this section we compare these collapse barriers with 
the linear overdensities of regions that
actually collapse to form halos in numerical simulations.

\begin{figure*}
\figurenum{2}
\epsscale{0.8}
\plotone{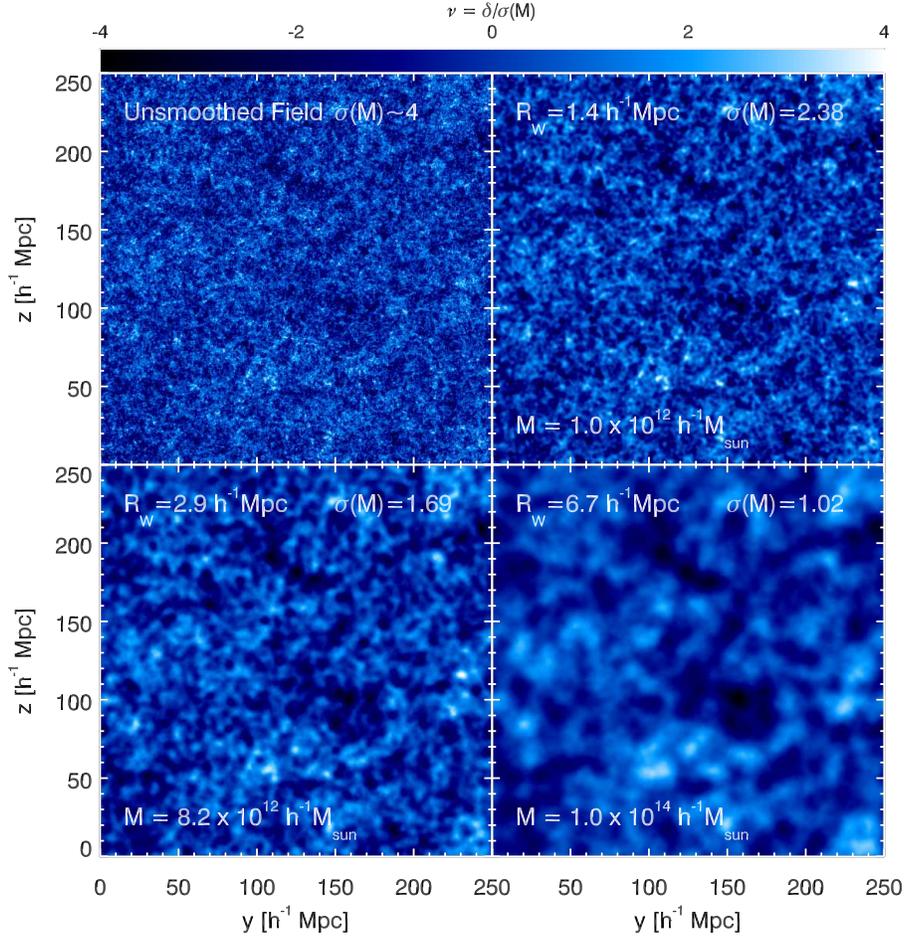}
\caption{\label{fig:smoothed_field}
Linear overdensity field as a function of smoothing scale for the L250 simulation at $a=0.01$. 
The upper left panel shows the linear
overdensity field interpolated onto a $512^{3}$ grid using the cloud-in-cell method (i.e., at full 
resolution of the simulation); the
effective smoothing of this grid corresponds to $\sigma(M)\sim4$. The other three panels show 
the field  $\delta(\vx,\Rw)$ smoothed on larger scales $\Rw$ using a real-space tophat window:
$\Rw=1.4h^{-1}\Mpc$ 
[$\sigma(M)=2.38, M=1.0\times10^{12}h^{-1}$],
$\Rw=2.9h^{-1}\Mpc$ 
[$\sigma(M)=1.69, M=8.2\times10^{12}h^{-1}\approx \Mstar$],
and $\Rw=6.7h^{-1}\Mpc$ 
[$\sigma(M)=1.02, M=1.0\times10^{14}h^{-1}$].
}
\end{figure*}

To calculate the linear overdensity of regions that later collapse to
form dark matter halos, we utilize cosmological simulations performed
with the Adaptive Refinement Tree code \citep[ART][]{kravtsov1997a}.  One
$512^{3}$ particle simulation models the formation of structure in the
\it Wilkinson Microwave Anisotropy Probe \rm (WMAP) 1st-year cosmology
\citep[][$\Omega_{m}=0.3$, $\Omega_{\Lambda}=0.7$, $\Omega_{b}=0.04$,
$n=1$, $\sigma_{8}=0.9$, $h=0.7$]{spergel2003a} in a cubic volume
$L=250~h^{-1}\Mpc$ on a side (hereafter, the L250 box).  This
simulation has a gravitational force resolution at the highest level
of refinement of $\epsilon_{\mathrm{L250}}=7.6~h^{-1}\kpc$, and a
particle mass of
$m_{p,\mathrm{L250}}=9.69\times10^{9}~h^{-1}M_{\sun}$.  An additional
$1024^3$ particle simulation of the WMAP 3rd-year cosmology
\citep[][$\Omega_{m}=0.27$, $\Omega_{\Lambda}=0.73$,
$\Omega_{b}=0.047$, $n=0.95$, $\sigma_{8}=0.79$,
$h=0.7$]{spergel2007a} with a larger volume ($L=1000~h^{-1}\Mpc$;
L1000W) is used to probe rare objects. This larger-volume simulation
has a comparably coarser resolution
($\epsilon_{\mathrm{1000W}}=30~h^{-1}\kpc$,
$m_{p,\mathrm{L1000W}}=6.98\times10^{10}~h^{-1}M_{\sun}$) than the
smaller simulation.

The simulations analyzed in this work were recently used by \cite{tinker2008a} as part of their
study of the universality of the halo mass function, and we use their 
halo catalogues when calculating the properties of the dark matter halo population.
\cite{tinker2008a} identified dark matter halos using a modified
spherical overdensity algorithm \citep[e.g.,][]{lacey1994a}, as
detailed in their \S 2.2.  Halo membership 
at a given redshift was determined by identifying 
peaks in the density field and assigning 
dark matter particles to peaks until the maximum
radius $R_{\Delta}$ of each halo contains a mean physical density
of $\Delta = M/[4\pi \rhobarm(z) R_{\Delta}^{3}/3]$, where $M$ is
the sum of the particle masses.  Results for other halo mass definitions are
examined in Appendix \ref{section:mass_definitions}.

Regions in the linear density field that collapse to form
halos are selected by identifying particles from the
$z=0$ halo catalogue in the simulation volume at early times.
For each simulation volume, we use the \cite{zeldovich1970a} 
approximation to re-scale the density field at 
the initial epoch of the simulation ($z>50$) to a sufficiently early epoch
at which the density field can be safely considered to be linear. 
We choose to re-scale all simulations to the scale factor $a=0.01$.
The density field for each simulation is calculated
by a cloud-in-cell interpolation of the particle distribution 
onto a $512^{3}$ grid.  This interpolated density field has
a mass resolution equal to the particle mass for the
L250 box, and eight times coarser mass resolution than
the L1000W box.  Hence, the field is
effectively smoothed on a scale of
$\sigma_{\mathrm{max}}(M)\approx4.2$ for L250 
and $\sigma_{\mathrm{max}}(M)\approx2.1$ for L1000W.
We will use each box only on scales $\sigma(M)<\sigma_{\mathrm{max}}(M)$.
Note that since we are interested in the regions that form
collapsed halos in the \cite{tinker2008a} catalogue, the largest $\sigma(M)$
of interest corresponds to the smallest halo for each simulation and 
all such regions are $\sigma(M)<\sigma_{\mathrm{max}}(M)$.

\subsection{Smoothing the Linear Overdensity Field}
\label{section:simulations:smoothed_field}

The linear overdensity field smoothed on a scale $\Rw$ can be calculated
directly from Equation~(\ref{equation:smoothed_overdensity}) by convolving
the density field with the window function $W(\vx,\Rw)$.  
A much more computationally-efficient
approach is to perform the convolution via multiplication in Fourier
space to obtain the transform of the smoothed overdensity field as
\begin{equation}
\label{equation:convolution}
\hat{\delta}_{\Rw}(k) = \hat{\delta}(k) \What(k,\Rw),
\end{equation}
\noindent
and then perform the inverse Fourier transform to arrive at the
smoothed overdensity $\delta_{\Rw}(\vx)$.
Here, $\hat{\delta}$ is the Fourier transform of the
unsmoothed density field and $\What(k,\Rw)$ is the
transform of the window function [for instance, $\What(k,\Rw)=\Whatr$
for a real-space tophat, see Equation~(\ref{equation:filter_real_space_tophat_transform})].
For each simulation volume, we compute this convolution 
for 150 smoothing scales from $\Rw\approx L/10$ to $\Rw\approx L/256$.
This results in $512^{3} \approx 134$~million
overdensity trajectories with 150 steps in $\sigma(M)$, as the change of overdensity
with decreasing smoothing scale at the location of each of the grid cells
is equivalent to integrating Equation~(\ref{equation:langevin}) 
with correlated large-scale modes.
We have verified that the root-mean-squared overdensity
fluctuations in each box are
$\langle\delta^{2}_{\Rw}(\vx)\rangle=\sigma^{2}(\Rw)$ 
simply by averaging over the gridded density field, 
in concordance with Equation~(\ref{equation:sigma}).

\begin{figure*}
\figurenum{3}
\epsscale{0.8}
\plotone{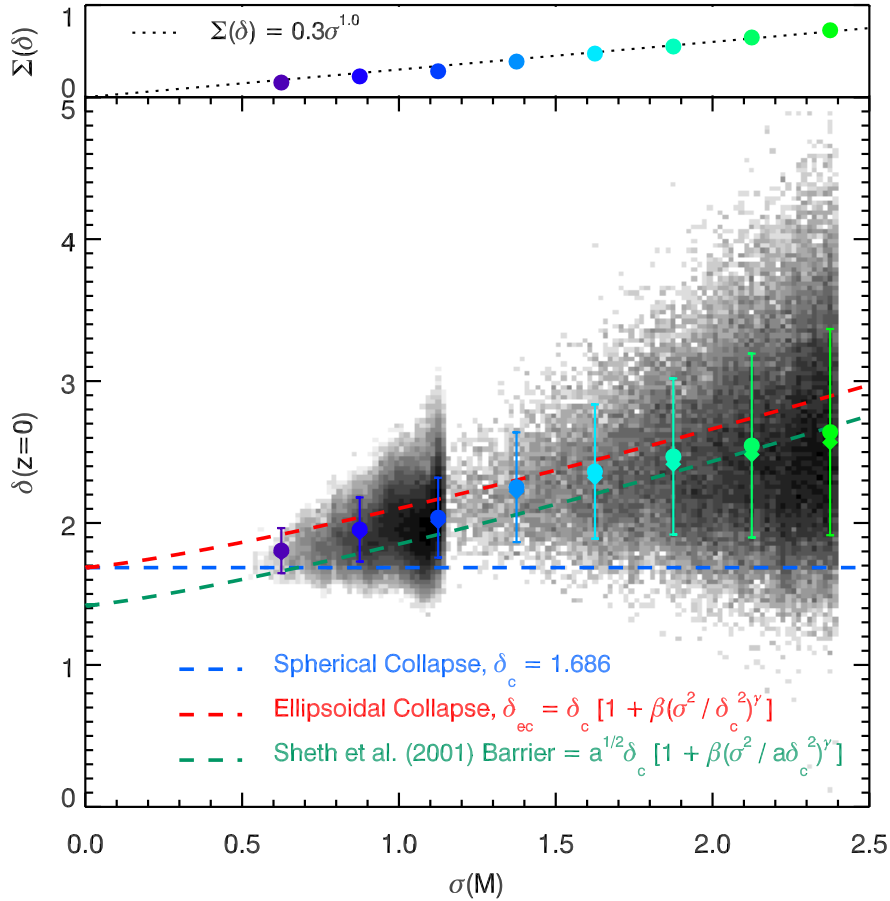}
\caption{\label{fig:smoothed_overdensity}
Smoothed linear overdensity $\delta$, extrapolated to $z=0$,  as a function of smoothing scale $\sigma(M)$ for regions
that collapse to form halos by $z=0$.   The circles 
correspond to the mean overdensities and the diamonds show the median overdensities, while the 
errorbars indicate the halo-to-halo scatter.  The error on the mean is significantly smaller than the 
scatter in all cases.  Shown
for comparison are the spherical collapse barrier ($\deltac$, blue dashed line), the \cite{sheth2001a}
ellipsoidal collapse barrier
($\deltaec$, red dashed line), and the collapse barrier associated with the \cite{sheth1999a} mass function
(green dashed line). 
The upper panel shows the variation in the scatter of barrier heights, where he have used the Greek letter 
``$\Sigma$'' to denote this scatter [not to be confused with $\sigma(M)$]. 
\\\\\\\\
}
\end{figure*}

Figure~\ref{fig:smoothed_field} shows the linear overdensity field
$\delta(a=1)=\delta(a=0.01)D(a=1)/D(a=0.01)$ of a thin slice through
the L250 simulation volume.  Shown are the unsmoothed field (upper
left panel) and the field smoothed on scales of $\sigma(M)=2.38$
($\Rw=1.4h^{-1}\Mpc$, $M\approx10^{12}h^{-1}\Msun$, approximately the
mass of the Milky Way halo, upper right panel), $\sigma(M)=1.69$
($\Rw=2.9h^{-1}\Mpc$, $M=M_{\star}=8.2\times10^{12}h^{-1}\Msun$, the
present collapse mass scale, lower left panel), and $\sigma(M)=1.02$
($\Rw=6.7h^{-1}\Mpc$, $M\approx10^{14}h^{-1}\Msun$, the mass of a
large group, lower right panel) using a real-space tophat filter
[Equation~(\ref{equation:filter_real_space_tophat})].  The figure
illustrates a variety of properties of the overdensity distribution
and the filter function as the smoothing scale is varied.  The
real-space tophat filter is broad in Fourier space, so the variations
in overdensities across many intervals in the smoothing scale are
correlated.  A sharp $k$-space filter would tend to decorrelate the
overdensities on small smoothing scales from larger scales as
independent frequency modes are added with increasing $\sigma(M)$.
The largest fluctuations in the unsmoothed overdensity field are
identifiable across the smoothed fields, reflecting the relation
between the initial density fluctuations and the eventual formation of
massive structures.

\begin{figure*}
\figurenum{4}
\epsscale{1.0}
\plotone{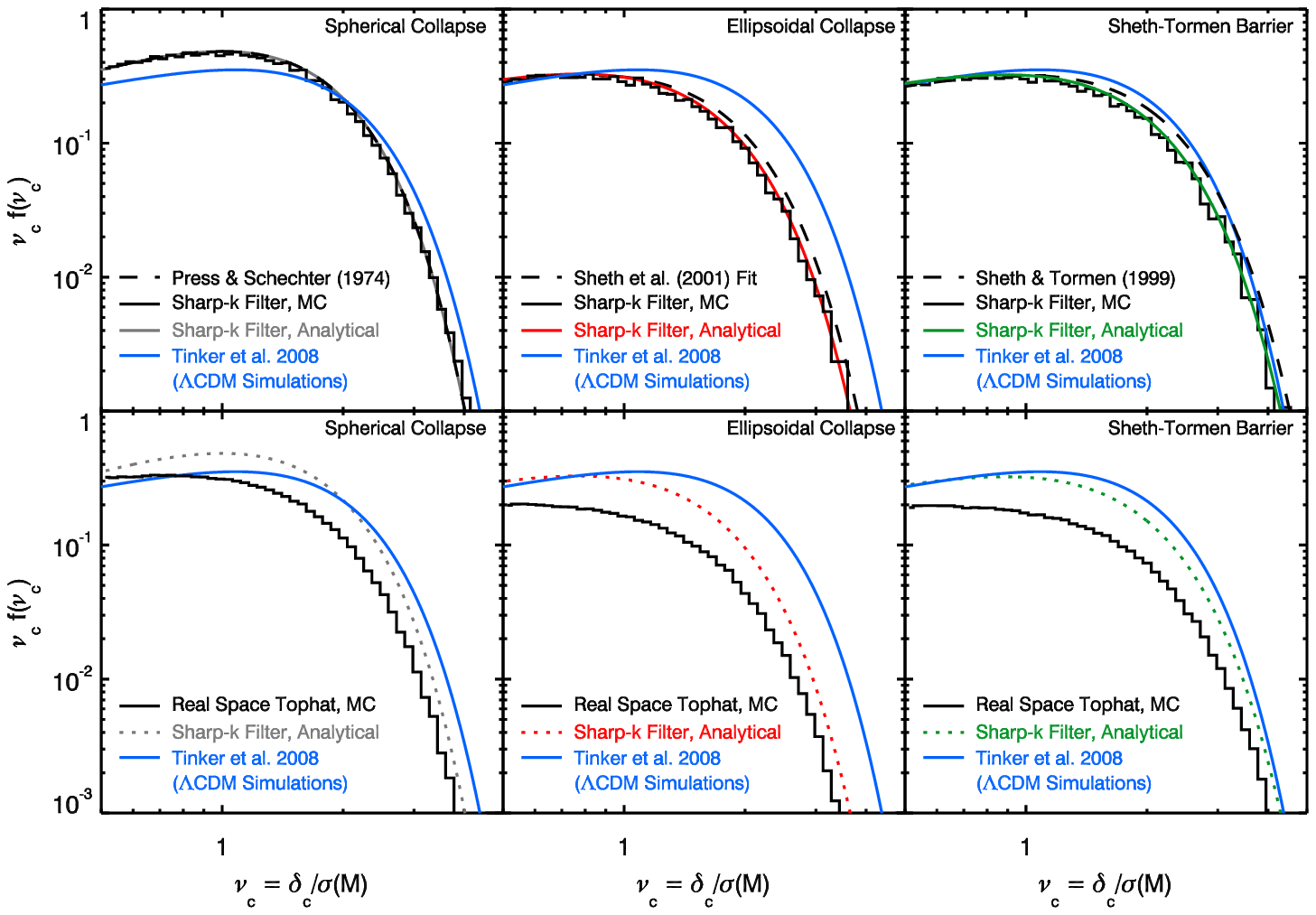}
\caption{\label{fig:fcds}
First-crossing distributions for common collapse barriers from the literature.
The upper row shows distributions calculated with the sharp $k$-space tophat filter
for 
the spherical collapse barrier 
\citep[][gray line, upper left panel]{gunn1972a},
the ellipsoidal collapse barrier 
\citep[][red line, upper middle panel]{sheth2001a},
and the collapse barrier associated with the \cite{sheth1999a}
mass function
\citep[][green line, upper right panel]{sheth2001a}.
The colored lines indicate the sharp $k$-space excursion set
first-crossing distributions calculated using the \cite{zhang2006a}
analytical method.  The histograms indicate Monte Carlo realizations
of the first-crossing distributions calculated by integrating the
Langevin Equation (\ref{equation:langevin}).  The dashed black lines
indicate the analytical form for the \cite{press1974a} mass function
(upper left panel), the \cite{sheth2001a} fit to the ellipsoidal
collapse first-crossing distribution (upper middle panel), and
the \cite{sheth1999a} mass function (upper right panel).  Also shown
is the \cite{tinker2008a} mass function determined from a large suite
of cosmological simulations (blue line, all panels).
In all cases, the Monte Carlo and \cite{zhang2006a} first-crossing
distributions agree well.  Since the ellipsoidal collapse
barrier lies above the spherical collapse barrier (see Figure \ref{fig:barriers}),
the corresponding first-crossing distribution predicts fewer
galaxy mass ($\nuc\gtrsim0.5$) halos than does the \cite{press1974a} formula.
The \cite{sheth1999a} mass function produces more high-mass halos by lowering
the associated collapse barrier below the spherical collapse barrier $\Bsc=\deltac$
at small $\sigma(M)$ (see Figure \ref{fig:barriers}).
For comparison, the bottom row shows the first-crossing distributions for the same barriers
calculated using a real-space tophat filter (histograms, bottom row).  In each case, using a real-space
tophat filter produces fewer halos in the excursion set calculation than does the sharp
$k$-space tophat for most halo masses.
\\\\\\
}
\end{figure*}

\subsection{Smoothed Overdensities of Collapsed Regions at $z=0$}
\label{section:simulations:smoothed_overdensity}

We can use the smoothed density fields to 
connect the final mass of a collapsed region [and
its associated fluctuation scale $\sigma(M)$] with
its initial smoothed overdensity linearly extrapolated to the
epoch of observation. 
For each halo identified in our catalgoues at $z=0$, 
we calculate the center-of-mass
of the halo particles from their positions in the linear 
density field at $a=0.01$ 
and use the window-smoothed field to compute the overdensity 
within the lagrangian 
radius\footnote{Although this is not entirely self-consistent 
(the smoothing scale is related to the mass scale by $M\propto (1+\delta)R^3$), 
it is clear that the error is small if this calculation is done
at an epoch when $\delta$ is very small. We have tested that the 
epoch we use $a=0.01$ is sufficiently early for this purpose.} 
$R = [3M/4\pi\rhobarm]^{1/3}$
about this location.
This overdensity
is then linearly extrapolated to $z=0$ to serve as an estimate of $\delta_{M}(\vx)$. 
We have checked that all of our conclusions are robust to specific choices 
regarding the smoothing procedure, such as the choice of initial positions (the $\vx$) 
of halos in the initial density field and the range of smoothing scales.

Figure~\ref{fig:smoothed_overdensity} shows the distribution of such
smoothed linear overdensities extrapolated to $z=0$ as a function of $\sigma(M)$.
The shaded regions represent the probability distribution
$p(\delta,\sigma)$ for regions that collapse to form halos defined 
relative to a $\Delta=200$ spherical overdensity identified in the
L1000W (601,448 halos) and  L250 (73,720 halos) boxes.
The median (colored diamonds) and mean
(colored circles) of $\delta(z=0)$ in bins of width
$\Delta\sigma(M)=0.25$ are measured for the halo population and shown
for comparison.  The distribution at fixed $\sigma(M)$,
$p(\delta|\sigma)$, is approximately log-normal in shape, with a width
that scales as $\Sigma(\delta) \approx 0.3\sigma^{1.0}(M)$ (indicated
by the error bars in Fig.~\ref{fig:smoothed_overdensity}).  
The scatter in $\delta(z=0)$ at fixed $\sigma(M)$
reflects both the intrinsic scatter in the linear overdensity of
collapsed regions and the limitations of our method to measure
$\delta(z=0)$ reliably for any individual halo.  The error on the mean
or median in any $\sigma(M)$-bin is much smaller than
$\Sigma(\delta)$.

At all measured halo masses, the mean and median linear overdensity of
the halo population at $\delta(z=0)$ exceeds the spherical collapse
overdensity $\deltac$. The dependence of the mean and median
overdensity on $\sigma(M)$ measured in simulations increases in a manner
that resembles the functional form of the
ellipsoidal collapse barrier $\deltaec$ 
[Equation~(\ref{equation:ellipsoidal_collapse_barrier})] and 
modified Sheth et al.~barrier $\deltaSMT$, but with a different normalization.  
Importantly, at the lowest values of variance (largest masses)
probed by the simulations [$\sigma(M)\sim0.5$] \it the measured overdensities of the
collapsed objects are larger than both the modified Sheth et al.~barrier $\deltaSMT$ 
and the spherical collapse barrier $\deltac$\rm.
We note that while our $L=1h^{-1}\Gpc$ $\LCDM$ simulations do not probe the 
highest masses ($M\gtrsim 5\times 10^{15}\, M_{\odot}$) 
and the  the largest scales ($\sigma(M)\lesssim0.4$) imaginable, 
in Appendix \ref{section:mass_definitions} we show that the conclusions developed
from these results do not change
even if the characteristic overdensity of regions that collapse to form halos
does not asymptote exactly to $\deltac$ for $\sigma\to0$.
In the next section of the paper, we explore the implication of these results
for halo mass functions calculated using the excursion set formalism.

\section{Excursion Set First-Crossing Distributions and Mass Functions in Cosmological Simulations}
\label{section:fcds}

The results presented in the previous section indicate that the
linearly-extrapolated overdensities of regions that collapse in
cosmological simulations behave in a manner analogous to the expectations of
the ellipsoidal collapse model. 
In this section, we
test the second ingredient of the excursion set {\it ansatz}: the
calculation of the first-crossing distribution and associated halo
mass function with a particular barrier.

\cite{tinker2008a} used a large suite of cosmological simulations to
determine an accurate numerical fit to the abundance of dark matter
halos as a function of their mass.  They found that the first-crossing
distribution that corresponds to the halo mass function measured
in simulations 
can be well-described by the function
\begin{equation}
\label{equation:tinker_fcd}
\nuc \fT(\nuc) = \AT\left[\left(\frac{\eT\nuc}{\deltac}\right)^{\dT} + \left(\frac{\nuc}{\deltac}\right)^{\gT}\right]\exp\left(-\frac{\hT\nuc^{2}}{\deltac^{2}}\right).
\end{equation}
\noindent
For $\Delta=200$ spherical overdensity halos, the best fit mas function
parameters are $\AT=0.482$, $\dT=1.97$, $\eT=1$, $\gT=0.51$, and $\hT=1.228$
\cite[with $\chi^{2}/\nu=1.14$, see table C4 of][]{tinker2008a}.

Figure~\ref{fig:fcds} compares the first-crossing
distribution given by Equation~(\ref{equation:tinker_fcd}) with 
first-crossing distributions calculated from
the excursion set formalism with the collapse barriers
described in \S~\ref{section:theory}. Where possible (i.e., in the case of 
sharp $k$-space tophat window function) we have checked our
Monte Carlo calculation of the first-crossing distribution
against the direct solution of the Volterra equation
using the method of \citet[][see Appendix \ref{section:zh06}]{zhang2006a}
and have found excellent agreement. 

The spherical collapse model (left column), corresponding to the
\cite{press1974a} mass function, displays the well-known deficit of
massive halos and overabundance of low-mass halos compared with
simulations (see the discussion in \S~\ref{section:theory:spherical_collapse_barrier} 
and Appendix~\ref{section:spherical_collapse}).  As demonstrated first by
\citet[][see their Figure 5]{bond1991a}, changing the window function
from the sharp $k$-filter to a real-space tophat filter  acts to
reduce the abundance of galaxy-mass halos rather than improve the
agreement between the spherical collapse and simulation first-crossing
distributions.

\begin{figure}
\figurenum{5}
\epsscale{1.0}
\plotone{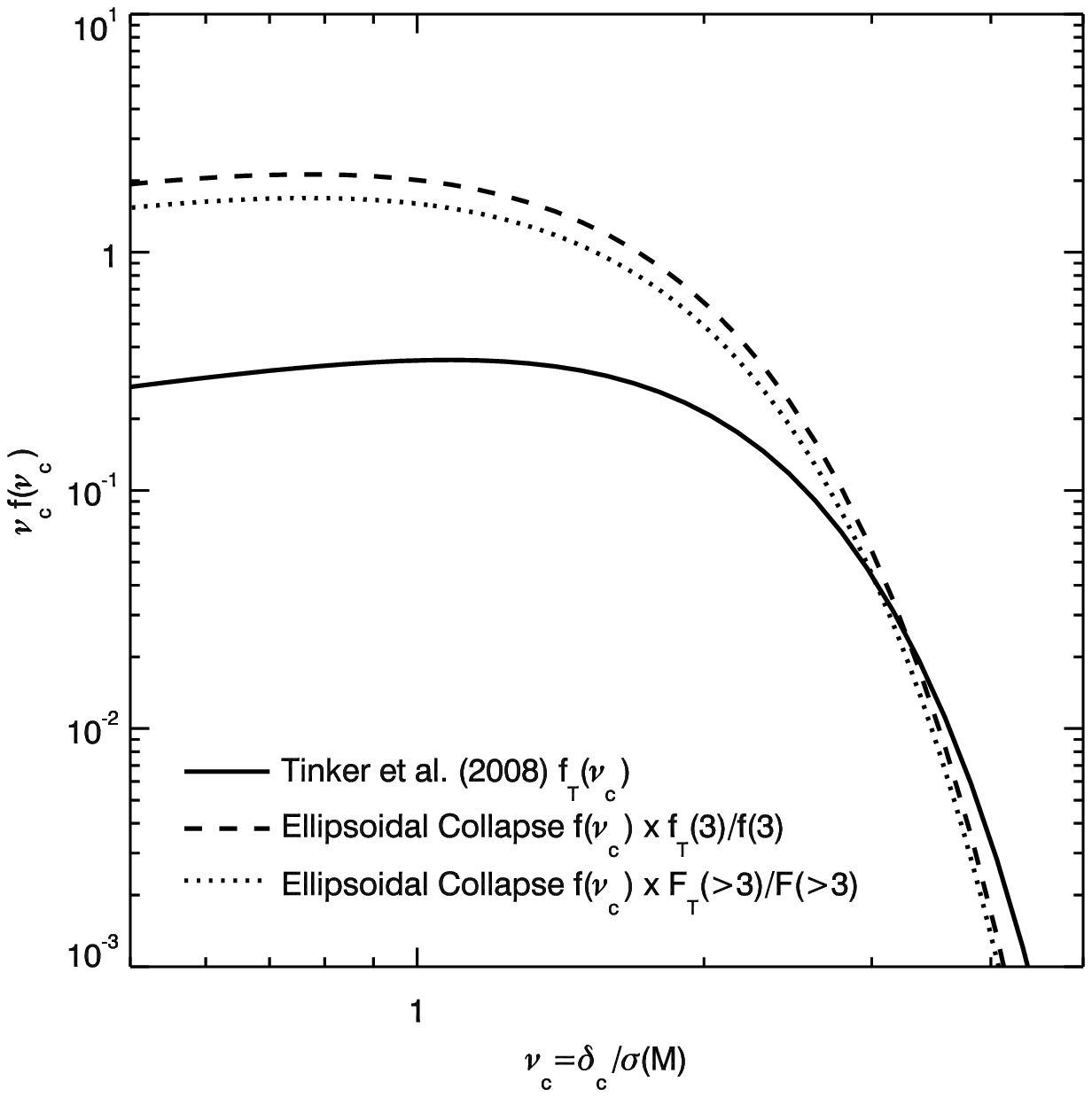}
\caption{\label{fig:fcd_norm}
Renormalized first-crossing distributions for ellipsoidal collapse vs. the \cite{tinker2008a} simulation results.
The form of the \cite{tinker2008a} mass function is normalized such that $\int \fT(\nuc)\dd \nuc=1$, but the
simulation results only probe masses $\nuc\gtrsim0.5$ that contain roughly $60\%$ the mass of the universe.
This figure demonstrates that any associated uncertainty with the normalization of the \cite{tinker2008a} mass
function (solid line) cannot reconcile the difference with the 
ellipsoidal collapse first-crossing distribution.  If the ellipsoidal collapse barrier 
(Equation \ref{equation:ellipsoidal_collapse_barrier})
first-crossing distribution is
renormalized to match the \cite{tinker2008a} mass function at $\nuc=3$ (dashed line), or renormalized to match
the integral of the \cite{tinker2008a} mass function at $\nuc>3$ (dotted line), the resulting distribution is
still discrepant from the \cite{tinker2008a} results.  Hence, the shape and normalization of the ellipsoidal
collapse barrier excursion set mass function differs from the halo mass function determined by N-body cosmological
simulations.
\\\\\\
}
\end{figure}

Using the ellipsoidal collapse barrier presented by 
SMT01 (see Equation \ref{equation:ellipsoidal_collapse_barrier})
with the excursion set calculation results in a lower first-crossing distribution at galaxy masses
(Figure~\ref{fig:fcds}, middle column) compared to the spherical collapse model.  These first-crossing 
distributions lie slightly below the 
SMT01 fitting function 
[Equation~(\ref{equation:ellipsoidal_collapse_fcd})], and
well below the \cite{tinker2008a} simulation results
for $\nuc>1$.  The ellipsoidal collapse barrier
converges to the spherical collapse barrier as 
$\sigma(M)\to0$, but at the mass scales  probed in
simulations 
[$\sigma(M)\gtrsim0.5$, $M\lesssim 10^{15}h^{-1}\Msun$] 
the ellipsoidal collapse barrier is considerably larger ($\deltaec\gtrsim1.8$).
The larger barrier height tends to suppress the
abundance of halos in this mass range as calculated
by the excursion set formalism.  Similar results
have been obtained by other authors 
\citep[e.g.,][their Figure 3]{sandvik2007a}, 
but the convergence to the spherical collapse
model at $\nuc>>1$ is seldom commented upon.
For consistency between the treatments of spherical
and ellipsoidal collapse, the abundance of halos
calculated for small ellipticities and prolaticities
(i.e., at large masses) must be
the same in both models.  As with the spherical
collapse model, the ellipsoidal collapse excursion
set mass function calculated with a real-space tophat
filter acts to lower the abundance of galaxy-mass
halos (Figure, \ref{fig:fcds}, bottom, middle panel).

The Sheth et al.~barrier 
[Equation~(\ref{equation:gif_simulation_barrier})], which SMT01
presented as the barrier corresponding to the GIF simulation (ST99) mass
function, produces a first-crossing distribution that lies below the
\cite{tinker2008a} simulation results (Figure~\ref{fig:fcds}, right
column). Note also that the predicted distribution does not 
agree with the SMT01
first-crossing distribution 
(Equation~(\ref{equation:sheth_tormen_fcd}), dashed line, Figure~\ref{fig:fcds},
upper right panel), which this barrier is designed to describe.  The
relative disagreement between the \cite{tinker2008a} and
SMT01 first-crossing distributions should be noted. This
disagreement, although relatively small on the scale of this figure,
is significant in terms of the halo abundance
\citep[see][]{tinker2008a}.  For the first-crossing distribution
determined using a real-space tophat window function (Figure
\ref{fig:fcds}, lower right panel), the Monte Carlo calculation lies
below the sharp $k$-space window function results and is discrepant with
the simulation result.

Since all simulated mass
functions are limited by their particle resolution one might
wonder if the differences with the simulations owe to 
the differences in normalization, as all of the excursion set
first-crossing distributions are normalized such
that $\int f(\nuc)\dd\nuc=1$.  The form of the \cite{tinker2008a}
first-crossing distribution [Equation~(\ref{equation:tinker_fcd})]
that is plotted in Figure~\ref{fig:fcds} is constructed to
require that $\int \fT(\nuc)\dd\nuc=1$.  However, the free parameters
of this function are fitted in a regime that incorporates only $\approx60\%$ of
the available mass.  To address this question, Figure~\ref{fig:fcd_norm} shows the 
ellipsoidal collapse first-crossing distribution
normalized at a fixed mass scale (instead of the integral constraint) to match the 
\cite{tinker2008a} first-crossing distribution at large masses ($\nuc=3$), or constrained
to provide the same integral over the same mass range (at $\nuc\ge3$).
The functions clearly differ in shape and not simply in normalization. 

Of the three barriers we examined, the excursion set
first-crossing distributions calculated from the
Sheth et al.~barrier most closely approximate the
\cite{tinker2008a} simulation results.  This result
is not surprising, given
that the ST99
function is itself a fit
to cosmological simulations. The
ellipsoidal collapse barrier produces a 
first-crossing distribution that differs substantially
from the ST99
fitting function, even though 
this mass function is frequently associated with 
ellipsoidal collapse model.

The results presented in this section demonstrate that {\it excursion set 
predictions disagree with $\LCDM$ simulation mass function results for all of these 
barriers.} Given that the overdensities of the collapsed regions
behave similarly to the barrier shape expected for the ellipsoidal
collapse, our results imply a manifest failure of the excursion set {\it ansatz}
as a method of computing the abundance of collapsed objects.   
In Appendix \ref{section:mass_definitions}, we show that our results hold for
other halo definitions, including Friends-of-Friends halos and other spherical
overdensity definitions.

\section{Discussion}
\label{section:discussion}

In the excursion set theory, the abundance, formation time, and bias of dark matter halos 
are assumed to be directly linked to the initial linear overdensity field extrapolated to a
given epoch using linear growth rates.  
This assumption significantly simplifies modeling, but clearly needs to be 
tested against direct numerical simulations. Our results show that 
the linear overdensities around the Lagrangian positions of 
the centers-of-mass of collapsed halos in simulations behave
in a manner analogous to the 
collapse overdensity barrier predicted by SMT01 using an ellipsoidal 
collapse model. The similar behavior
suggests that this model captures the main physics behind the nonlinear gravitational
collapse around peaks in the initial density field.  

At the same time, the failure of the excursion set {\it ansatz} to
predict correct abundance of collapsed halos with the ellipsoidal
collapse barrier demonstrates that the {\it ansatz} is flawed. Several
aspects of the {\it ansatz} may be responsible for its failure to
accurately describe simulation results.  For example, assumption that
each mass element can be assigned to a collapsed halo using only the
local overdensity independently of its environment is definitely
problematic. The threshold overdensity for collapse (the barrier) in
either the spherical or ellipsoidal collapse models is predicted for
the volumes centered on a density peak, and not for a random mass
element in the field.  While the actual collapse will occur around the
density peaks, additional mass near a peak may collapse onto it even
though the extended region may not satisfy the local collapse
condition. 
Our study highlights the failure of the excursion set {\it ansatz}
on a statistical basis for the entire halo population, but a physical
model for the collapse and formation of dark matter halos must also
succeed object-by-object \citep[see, e.g.,][]{katz1993a}.
The methods for calculating the abundance of halos that
treat individual peaks in the density field \citep[][Dalal et al., in
prep.]{bond1996a,monaco2002a} may therefore afford a better way to predict
the formation, masses, and abundances of collapsed objects.  

The shape of the effective collapse barrier is determined both by the distribution of
overdensities in a smoothed Gaussian field (e.g., with ellipticities and
prolaticities, or other properties, sufficient for collapse) and by the
dependence of the collapse condition on the overdensity (and/or other properties).
The halo formation time can be defined relative to the mean assembly history
\citep[e.g.,][]{wechsler2002a}, but the specific choice of definition can
influence the relation between formation time and halo mass 
\citep[for a recent discussion, see][]{li2008a}.
In terms of the linear overdensity and the effective barrier $B$, one definition of 
halo collapse is simply that of Eq.~(\ref{equation:collapse_barrier}).  Of course,
determining the effective barrier $B$ or measuring the collapse epoch $z_{c}$
complicates the matter.  
The definition of collapse can be connected to the physical properties of
halos through dynamical models, but identifying evolutionary phases 
in such simple models with the actual nonlinear growth of dark matter
halos may be incorrect or inaccurate.
For instance, the ellipsoidal collapse model of \cite{bond1996a}
associates halo formation with the collapse of the longest ellipsoid axis, while freezing
the collapse of the shorter axis at a particular point to prevent a density singularity. 
Halo virialization, however, may be associated with the collapse of another axis or
conditional properties of shape of the density or shear fields \citep[e.g.,][]{lee1998a,monaco2002a}.
These issues warrant a more careful examination that we defer for future work now that we 
have a statement of the problems at hand.

Dark matter halos reside in special locations of the density field, and
have different clustering properties than bulk matter \citep[e.g.,][]{kaiser1984a,efstathiou1988a,cole1989a}.
Models of halo bias are tightly connected with the effective collapse barrier.
\cite{mo1996a} presented the idea that the biasing of halos is connected
to the rate at which density trajectories cross two separate barriers,
and used this concept to calculate the Lagrangian bias of halos in the
Press-Schechter model.  
SMT01 adapted these ideas to calculate
the halo bias implied by their modified ellipsoidal collapse model.  

In principle, the
same collapse barrier should predict consistently both the bias and
mass function of halos.  However, while the 
ST99 mass
function models the halo abundance reasonably well, \cite{seljak2004a}
found that the 
SMT01 bias model does not work well at low
masses.  For massive halos at high $\nuc$, \cite{cohn2008a} found that
the \cite{mo1996a} bias scaling works better than the
SMT01 model \citep[however, see][]{hu2003a}.
Interestingly, \cite{reed2008a} found that early-forming halos in
their simulations were well-modeled by 
SMT01 at peak
heights as large as $\nuc\sim4$ (at small $\nuc$, the large-scale bias
they measure drops below the SMT01
model, in a manner
similar to that found by \citealt{seljak2004a}).  We note that if the
effective collapse barrier converges to the spherical collapse
overdensity $\deltac$ for large mass or highly-biased halos, then one
might expect the halo bias to mimic the \cite{mo1996a} scaling at
large $\nuc$.  We speculate that the intriguing deviations between the
models for halo bias and abundance of halos are connected to the
current discrepancy between the effective collapse barrier and the
abundance of halos in the excursion set {\it ansatz} that we discuss
in this paper.

Lastly, the effective collapse barrier may be connected with the
assembly bias phenomenon where dark matter halo clustering correlates
with formation time.  \cite{gao2005a} found that low-mass halos 
($M<\Mstar$) that formed early were much more strongly clustered than
late-forming halos of the same mass \citep[see also][]{sheth2004a,harker2006a}. \cite{wechsler2006a} demonstrated
that the sense of assembly bias reverses at high-mass ($M>\Mstar$) such
that late-forming halos were more strongly clustered than early-forming
halos at fixed mass, and that these correlations are reflected in the
relative bias of halos with different concentrations at fixed mass \cite[see also][]{wetzel2007a,jing2007a}.

\citet{zentner2007a} in his review showed 
that window functions that are local in real-space rather than Fourier 
space naturally result in early-forming halos that reside in underdense 
regions, as reported for high-mass halos ($M>\Mstar$) in the simulations.  
The standard implementation of the excursion set formalism assumes that 
the process of nonlinear collapse can be encapsulated into the assignment of 
the collapse barrier.  For high-mass halos, tidal influences and nonlinear interactions 
with nearby objects should be minimal because halos with masses significantly 
larger than $\Mstar$ form from nearly spherical peaks and 
usually dominate their local environments.  As a result, the excursion set assumption 
should be most valid for high-mass ($M \gg \Mstar$) halos and leads 
to a natural picture where early-forming, high-mass halos become less 
strongly clustered than their late-forming counterparts.

The reversal of the environment-dependent halo formation trend at low-mass 
may owe to the truncation of small halo growth by nearby structure as suggested by \citet{wang2007a}.  
\cite{dalal2008a} greatly extended this work, validating the high-mass trend in a set of scale-free 
numerical simulations and showing that environmental influences on halo 
bias, concentration, and formation time at fixed mass could be accounted 
for by considering the ``peak curvature'', $\dd \delta_{\Rw}/\dd \sigma$, in addition 
to peak height.  The peak curvature serves as some proxy for environment as peaks with 
greater curvature lie in relatively underdense environments.  
\cite{dalal2008a} also used a toy model to demonstrate that at small masses
environmental effects, such as those suggested by \cite{wang2007a}, 
can truncate halo growth and drive early-forming halos to become less anti-biased as
they are advected by the larger-scale matter field.

Our results relate to these findings by providing a new
outlook on the connection between halo abundance, the effective
collapse barrier, and the excursion set formalism.  If the excursion
set formalism fails to account properly for the abundance of
halos using an appropriate form for the collapse barrier, then it may also fail
to describe reliably the connection between formation time, mass,
and properties of the density field.  The effective collapse barrier
may well be a function of additional parameters beyond the
local density \citep[e.g.,][]{chiueh2001a,sandvik2007a}, 
may incorporate information about the larger-scale field 
\citep[e.g.,][]{zentner2007a,desjacques2008a,dalal2008a}, and 
may require additional parameter dependencies that to account for the 
nonlinear collapse of overdensities \citep[e.g.,][]{wang2007a,dalal2008a}.

At present, the excursion set theory provides the main framework used 
to develop heuristic understanding of simulation results and to formulate 
fits to simulation results for halo abundance, clustering, and other halo 
properties.  The method succeeds in a gross sense.  Excursion set theory identifies the 
fundamental scale in the problem, the mass where $\sigma(M) \sim \deltac$.  
Below this characteristic mass  
the halo abundance per unit mass has a simple, power-law form, while  above this mass the
halo abundance drops rapidly.
However, a precise understanding of halo abundance and clustering beyond
the gross accuracy provided by the excursion set {\it ansatz} is now necessary.
Contemporary and forthcoming efforts to use measurements of 
the abundance and clustering of galaxy clusters to constrain cosmological parameters,
as well as comprehensive statistical studies of galaxy formation and evolution, only 
highlight the need for a sound understanding of halo abundance and assembly, both globally and 
as a function of environment. We need to explore amendments and alternatives 
to the excursion set model to make progress in our understanding of halo formation.

\section{Summary}
\label{section:summary}

In this paper we have presented tests of the excursion set
{\it ansatz} against cosmological simulations.  Using a subset of
cosmological simulations from the \cite{tinker2008a} study of the halo
mass function, we identify the locations in the linear overdensity
field that later collapse to form dark matter halos.  
We demonstrate
that the dependence of the linear overdensity of 
these regions on mass or smoothing scale $\sigma(M)$
resembles predictions of the ellipsoidal collapse model. 
While the effective collapse barrier of simulated halos behaves analogously to 
the simple ellipsoidal collapse barrier,
the simulated halo mass function is inconsistent with what
the excursion set {\it ansatz} predicts for such
a barrier. This inconsistency implies that the excursion set {\it ansatz} is not valid 
and cannot be used reliably to predict halo abundance or bias.

The modified collapse barrier of \cite{sheth2001a} differs
significantly from the physical behavior calculated by the ellipsoidal
collapse model, which for example predicts convergence to the
spherical barrier ($\delta\to\deltac$) for the rarest peaks
($\sigma\to0$). In view of this, the interpretation of the
\cite{sheth1999a} mass function as a prediction of the ellipsoidal
collapse model for the abundance of dark matter halos is not correct.

The impressive statistics of ever-larger
dark matter simulations will likely continue to uncover
increasingly subtle variations on the classical
picture of dark matter halo formation.  In this work,
we identify a striking inconsistency between the effective
collapse barrier of simulated halos and excursion set
formalism predictions for their abundance. Our results also demonstrate
that there is still much to learn and understand about the conditions for 
and the process of halo collapse, which warrants further studies 
that critically revisit these issues using modern large cosmological
simulations.

\acknowledgments

We would like to thank Anatoly Klypin for access to his cosmological simulations
and 
Neal Dalal for useful discussions on the
subject of this study. BER gratefully acknowledges support from a
Spitzer Fellowship through a NASA grant administrated by the Spitzer
Science Center. AVK is supported by the NSF under grants AST-0239759
and AST-0507666 and by NASA through grant NAG5-13274. BER and AVK are
also partially supported by the Kavli Institute for Cosmological
Physics at the University of Chicago. ARZ is supported by the
University of Pittsburgh and by the NSF through grant AST 0806367.
ARZ would like to thank the Michigan Center for Theoretical Physics at
the University of Michigan for support and hospitality while some of
this work was being performed.  Some of the calculations used in this
work have been performed on the Joint Fermilab - KICP Supercomputing
Cluster, supported by grants from Fermilab, the Kavli Institute for
Cosmological Physics, and the University of Chicago. One of the
simulations was performed at the Leibniz Rechenzentrum Munich, partly
using German Grid infrastructure provided by AstroGrid-D.

\appendix
\section{Spherical Collapse}
\label{section:spherical_collapse}

A simple approximation for the dynamical
evolution of an overdense region is
the spherical collapse model.
The dynamical equation for the expansion and
collapse of a spherical overdensity $\Delta$ of physical size $R_{\mathrm{phys}}(z)$
in a flat universe with a cosmological constant 
can be written \citep[e.g.,][]{gunn1972a}
\begin{equation}
\label{equation:spherical_collapse}
\frac{\dd^{2}\aR}{\dd t^{2}} = \frac{8}{3}\pi G \rhobarl \aR - \frac{4}{3}\pi G\rhobarm \aR [1+\Delta(t)],
\end{equation}
\noindent
where $\aR = R_{\mathrm{phys}}/R$ is the scale factor of the
region, $\rhobarl=\Omega_{\Lambda}\rho_{c}$ is the dark energy
density, and $\Omega_{\Lambda}$ is the dark energy density
parameter.
The corresponding growing {\it linear} overdensity $\delta$ obeys the differential equation
\begin{equation}
\label{equation:linear_growth_equation}
\frac{\dd^{2}\delta}{\dd t^{2}} + 2H\frac{\dd \delta}{\dd t} = 4\pi G\rhobarm \delta
\end{equation}
\cite[e.g.,][]{lifshitz1946a,peebles1965a}, while the physical overdensity evolves as
\begin{equation}
\label{equation:spherical_collapse_physical_overdensity}
\Delta(t) = \frac{a^{3}}{\aR^{3}}-1,
\end{equation}
\noindent
where $a$ is the universal scale factor.  
The initial conditions for evolving Equation \ref{equation:spherical_collapse}
are simply
\begin{equation}
\label{equation:spherical_collapse_initial_conditions_a}
\aR(\tinit)=a(\tinit)(1-\delta(\tinit)/3),
\end{equation}
\begin{equation}
\label{equation:spherical_collapse_initial_conditions_adot}
\dot{\aR}(\tinit)=H(\tinit)\aR(\tinit)-a(\tinit)H_{D}(\tinit)\delta/3,
\end{equation}
\noindent
where $H_{D}=\dot{D}/D$ describes the rate of change of overdensities.
Initially $\Delta(t)\simeq\delta(t)$, as can be checked by
Taylor-expanding Equation \ref{equation:spherical_collapse_physical_overdensity} 
and comparing it with Equation \ref{equation:spherical_collapse_initial_conditions_a}, but
eventually the quantities diverge as the overdensity begins to exceed the applicability
of the linear order approximation.  As the spherical region begins to break from
the universal expansion and reaches a maximum radius at the turn-around time $t=t_{\mathrm{ta}}$, the physical
overdensity reaches $1+\Delta(t=t_{\mathrm{ta}})\simeq5.55$ while the linear overdensity is 
$\delta(t=t_{\mathrm{ta}})\simeq1.06$.
As the region collapses to a point of zero size, $\Delta\to\infty$ while the linearly
extrapolated overdensity approaches a value of
\begin{equation}
\label{equation:spherical_collapse_barrier_appendix}
\Bsc \equiv \deltasc = \delta(\tinit)D(z_{c})/D(\zinit) = \delta_{c}\,\,\,\,\mathrm{(Spherical~Collapse),}
\end{equation}
where $\delta_{c}=1.686$ is often called the linear collapse overdensity
for the growth of spherical perturbations \citep{gunn1972a,peebles1980a}
for an $\Omega_{m}=1$ universe \citep[$\delta_{c}$ has a weak
dependence on cosmology, see, e.g.,][]{bond1996a,eke1996a}.
If the spherical collapse of the region to zero size is associated with
the formation of a dark matter halo, then 
Equation \ref{equation:spherical_collapse_barrier_appendix} can be used
as the ``spherical collapse barrier'' for purposes of calculating
the first-crossing distribution associated with spherical collapse.

\section{Ellipsoidal Collapse and Its Modifications}
\label{section:ellipsoidal_collapse}

The gravitational collapse calculation has been generalized 
to model nonspherical collapse by a number of
authors \citep[e.g.,][]{zeldovich1970a,nariai1972a,hoffman1986a,bertschinger1994a,eisenstein1995a,bond1996a,audit1997a,del_popolo2001a,shen2006a}.  Below we focus on an ellipsoidal collapse model by
\citet[][see their \S 2.1.3 and Appendix A]{bond1996a},
which treats the gravitational collapse of a
homogeneous ellipsoid
by separately following the coupled evolution 
of each axis $a_{i}$ of the ellipsoid.
\cite{bond1996a} showed that
the dynamical equation for the expansion and
collapse of an ellipsoidal region with linear overdensity $\delta(t=\tinit)=\delta_{0}$
at an initial time $t_{\mathrm{init}}$ can be written 
\begin{equation}
\label{equation:ellipsoidal_collapse}
\frac{\dd^{2}a_{i}}{\dd t^{2}} = \frac{8}{3}\pi G \rhobarl a_{i} - 4\pi G\rhobarm a_{i} \left[\frac{1}{3}+\frac{\Delta(t)}{3} + \frac{b'_{i}(t)}{2}\Delta(t) + \lambda_{i}'(t)\right]
\end{equation}
\noindent
where the index $i=1,2,3$ indicates a principal axis.
The term
\begin{equation}
\label{equation:triaxial_acceleration}
b'_{i} = -\frac{2}{3} + a_{1}a_{2}a_{3} \int_{0}^{\infty} \frac{\dd \tau}{(a_{i}^{2}+\tau)\Pi_{m=1}^{3} (a_{m}^{2}+\tau)^{1/2}}
\end{equation}
\noindent
describes triaxial contributions to the gravitational acceleration.
A linear order approximation of the effects of external tides on the evolution of the
region are included through the factors $\lambda'_{i}(t) = \lambda_{i} - \delta/3$,
which are written in terms of the eigenvalues of the strain tensor
\begin{equation}
\label{equation:strain_eigenvalue_3}
\lambda_{3} = \frac{\delta}{3}(1 + 3e + p),
\end{equation}
\begin{equation}
\label{equation:strain_eigenvalue_2}
\lambda_{2} = \frac{\delta}{3}(1 - 2p),
\end{equation}
\begin{equation}
\label{equation:strain_eigenvalue_1}
\lambda_{1} = \frac{\delta}{3}(1 - 3e + p).
\end{equation}
\noindent
Here, $e\ge0$ is the ellipticity, $-e\le p \le e$ is the prolaticity,
and the eigenvalues are ordered $\lambda_{3}\ge\lambda_{2}\ge\lambda_{1}$.
This linear order approximation to the effects of tides
grows as $\lambda_{i}(t)\propto\delta(t)\propto D(t)$.
The physical overdensity in the ellipsoidal collapse model evolves simply as
\begin{equation}
\label{equation:ellipsoidal_overdensity}
\Delta(t) = \frac{a^{3}}{a_{1}a_{2}a_{3}}-1,
\end{equation}
\noindent
where $a$ is the universal scale factor.
The initial conditions for evolving Equation \ref{equation:ellipsoidal_collapse}
are 
\begin{equation}
\label{equation:ellipsoidal_collapse_initial_conditions_a}
a_{i}(\tinit)=a(\tinit)[1-\lambda_{i}(\tinit)],
\end{equation}
\begin{equation}
\label{equation:ellipsoidal_collapse_initial_conditions_adot}
\dot{a}_{i}(\tinit)=H(\tinit)a_{i}(\tinit) - a(\tinit)H_{D}(\tinit)\lambda_{i}(\tinit),
\end{equation}
\noindent
The evolution of the region is then determined by the cosmology, the initial
overdensity $\delta_{0}$, the ellipticity $e$, the prolaticity $p$,
and the initial universal scale factor $\bar{a}(t=\tinit)$.  
For a spherical system ($e=0$, $p=0$, $a_{1}=a_{2}=a_{3}$), the dynamical 
equations reduce to Equation~(\ref{equation:spherical_collapse}).

\cite{sheth2001a} used the results of
\cite{doroshkevich1970a} to show that, in the
context of the \cite{bond1996a} ellipsoidal collapse
model, the 
formula
\begin{equation}
\label{equation:shape_distribution}
g(e,p|\delta) = \frac{1125}{\sqrt{10\pi}}e(e^2 - p^2)\left(\frac{\delta}{\sigma}\right)^5 \exp\left[-\frac{5}{2}\frac{\delta^{2}}{\sigma^{2}}(3e^{2} +p^{2})\right]
\end{equation}
provides the expected distribution 
of ellipticities and prolaticities
for the shear field of Gaussian random overdensities 
and
corresponds to a
distribution of effective collapse barriers for 
halos
 (as a function of
$e$ and $p$) that describes the overdensity
at which the
last (longest) principal axis collapses.
\cite{sheth2001a} provided an empirically-determined, implicit functional form to approximate the
shape of the ellipsoidal collapse barrier in terms
of $e$ and $p$,
\begin{equation}
\label{equation:ellipsoidal_collapse_implicit}
\frac{\deltaec(e,p)}{\deltasc} = 1 + \beta\left[5(e^{2}\pm p^2) \frac{\deltaec^{2}(e,p)}{\deltasc^{2}}\right]^{\gamma},
\end{equation}
\noindent
which follows the most probable ($p=0$) trend of the 
collapse barrier distribution well for the parameter values
$\beta=0.47$ and $\gamma=0.615$ 
\citep[a more recent calculation has found $\beta=0.412$ and $\gamma=0.618$,][]{desjacques2008a}.
For the most probable prolaticity $p=0$, the maximum of the
probability distribution $g(e,p=0|\delta)$
follows the ridgeline
$e_{\mathrm{mp}}=(\sigma/\delta)\sqrt{5}$.  Substituting
$p=0$ and $e=e_{\mathrm{mp}}$ into Equation \ref{equation:ellipsoidal_collapse_implicit}
yields a characteristic ellipsoidal collapse barrier in
terms of the overdensity variance as
\begin{equation}
\label{equation:ellipsoidal_collapse_barrier_appendix}
\Bec \equiv \deltaec = \deltasc \left[ 1 + \beta\left( \frac{\sigma^{2}}{\deltasc^{2}}\right)^{\gamma}\right].
\end{equation}
Note that this barrier reduces to the spherical collapse barrier 
[$B=\delta_{c}$, Equation~(\ref{equation:spherical_collapse_barrier})]
for large halo masses (small variances).
For reference, the ellipsoidal collapse barrier is plotted in 
Figure \ref{fig:barriers} with the \cite{sheth2001a} parameters (dashed
red line).

\section{An Integral Method for First-Crossing Distributions with Sharp $k$-Space Filtering}
\label{section:zh06}

Typically, the first-crossing distribution is determined
with a Monte Carlo approach (effectively by integrating 
Equation \ref{equation:langevin}).  For the case of sharp $k$-space
filtering, where smoothed overdensity executes a Markovian
random walk with the smoothing scale, and barriers with a
suitably weak dependence on the variance,
\cite{zhang2006a} showed that by properly accounting for the 
rate of first barrier crossings in an ensemble of trajectories,
an integral relation for the 
first-crossing distribution could be expressed in terms of
the functional form of the barrier $B$ and the probability
$P(\delta,S)$ that a trajectory $\delta$ 
first crosses the barrier near $S$.
Specifically, in terms of the barrier $B(S)$, the
first-crossing distribution satisfies
\begin{equation} 
\label{equation:zh06_integral_constraint}
1 = \int_{0}^{S} f(S')\dd S' + \int_{-\infty}^{B(S)}P(\delta,S)\dd\delta \,\,\,\,\,\,\,(\mathrm{sharp-}k~\mathrm{filtering}).
\end{equation} 
\noindent
The first term on the right hand
side accounts for trajectories that have crossed at 
scales larger than $S$, while 
the second term 
\begin{equation}
\label{equation:zh06_sharpk_crossing_probability}
P(\delta,S) = P_{0}(\delta,S) - \int_{0}^{S}\dd S' f(S')P_{0}(\delta-B(S'),S-S')
\end{equation}
subtracts the rate of down-crossings from the probability 
\begin{equation}
\label{equation:zh06_gaussian_crossing}
P_{0}(\delta,S) = \frac{1}{\sqrt{2\pi S}}\exp\left(-\frac{\delta^{2}}{2S}\right)
\end{equation}
\noindent
that a trajectory given crosses the barrier near $S$.
Equation \ref{equation:zh06_integral_constraint}
can be differentiated and combined with Equations 
\ref{equation:zh06_sharpk_crossing_probability}-\ref{equation:zh06_gaussian_crossing}
to produce a Volterra integral
equation of the second kind for the 
the first-crossing distribution 
\begin{equation}
\label{equation:zh06_volterra_fs}
f(S) = g_{1}(S) + \int_{0}^{S} f(S)g_{2}(S,S')\dd S' \,\,\,\,\,\,\,(\mathrm{sharp-}k~\mathrm{filtering}),
\end{equation}
\noindent
with
\begin{equation}
\label{equation:zh06_g1}
g_{1}(S) = \left[\frac{B(S)}{S} - 2 \frac{\dd B}{\dd S}\right]P_{0}(B(S),S)
\end{equation}
\begin{equation}
\label{equation:zh06_g2}
g_{2}(S,S') = \left[2\frac{\dd B}{\dd S} - \frac{B(S)-B(S')}{S-S'}\right]P_{0}(B(S)-B(S'),S-S')
\end{equation}
\noindent
Hence, given a barrier shape $B(S)$, \cite{zhang2006a} have
provided a helpful method for calculating the 
first-crossing distribution for a $k$-space filter 
via Equation \ref{equation:zh06_volterra_fs} .

\section{The Excursion Set {\it Ansatz } and Halo Mass Definitions}
\label{section:mass_definitions}

The results of \S \ref{section:simulations} demonstrate a disconnect between the characteristic
linear overdensity of regions that collapse to form dark matter halos and the collapse barrier
required to reproduce the abundance of those same halos using the excursion set formalism.  
While the results of \S \ref{section:simulations}
are internally consistent, 
one might wonder if the failure of the excursion set {\it ansatz} was peculiar to the $\Delta=200$ spherical
overdensity halo definition.
In this Appendix, we demonstrate that the excursion set {\it ansatz} also fails for other common
halo definitions (specifically, $\Delta=100$ spherical overdensity halos, 
$\Delta=600$ spherical overdensity halos, Friends-of-Friends
halos, and halos defined by spherical regions of size $R=2R_{200}$).  Since these
halo definitions span the most practical definitions found in the literature, the results of this appendix 
present an exhaustive study of how our results depend on the halo and mass definitions.  Further, since the
largest simulation we study has a $1h^{-3}\,\Gpc^{3}$ volume, the linear overdensity of regions with very large
mass ($M_{200}\gtrsim5\times10^{15}h^{-1}\Msun$) are not probed by our simulations.  Below, using
extrapolations of the $\delta-\sigma(M)$ trend for various halo mass definitions, we demonstrate that
even if $\delta\not\to\delta_{c}$ as $\sigma(M)\to0$ the excursion set mass functions do not reproduce the
simulated mass function for any halo mass definition we consider.  

To repeat the calculations in \S \ref{section:simulations} for other halo definitions, we must construct
additional halo catalogues.
For the spherical overdensity halo definition, we follow \cite{tinker2008a}
and define halos with an overdensity $\Delta$ relative to the background density $\rhobarm$ as the particles within 
a radius $R_{\Delta}$ around density peaks.  For the Friends-of-Friends halo definition
\citep[e.g.,][]{davis1985a}, we adopt the standard linking length of $b=0.2$.  For halos defined by
spherical regions of size $R=2R_{\Delta}$, we use the $\Delta=200$ catalogue to identify
halos and redefine the halo masses by assigning all particles within $2R_{200}$ of the
center-of-mass membership
in the halo.  If the radius $R = 2R_{200}$ for one halo includes the center-of-mass of a smaller halo, the smaller halo
is discarded from the catalogue.

For each halo definition, the mass function $\dd n/\dd M$ is determined by constructing a histogram
for the halos by binning in mass.  
The first-crossing
distributions $f(\nuc)$ corresponding to each mass function are calculated using Equation~(\ref{equation:mass_function}).
For each mass bin we calculate jack-knife errors, as described in detail by \cite{tinker2008a}.  
For two spherical overdensity definitions, $\Delta=200$ and $\Delta=600$, we simply adopt
a mass function 
of the form of Equation~(\ref{equation:tinker_fcd})
with the best-fit parameters determined in Appendix C of \cite{tinker2008a}.  
For $\Delta=200$, $\AT=0.482$, $\dT=1.97$, $\eT=1$, $\gT=0.51$, and $\hT=1.228$.
For $\Delta=600$, $\AT=0.494$, $\dT=2.56$, $\eT=0.93$, $\gT=0.45$, and $\hT=1.553$.
These analytical mass function fits make use of the wide range of simulations studied by \cite{tinker2008a}.
For mass functions for the other halo definitions, we rely on our halo catalogues for the
L1000W simulation and represent the mass function with binned values and uncertainty estimates constructed from
these catalogues.
We have checked that the first-crossing distribution calculated from the binned mass function
for the $\Delta=200$ and $\Delta=600$ halos in the L1000W simulation 
match the corresponding analytical fits from \cite{tinker2008a} extremely well, and we therefore
expect that the binned mass functions and first-crossing distributions for the other halo mass definitions
are reliable estimates of the halo abundance over the mass range probed by the L1000W simulation.
The L1000W simulation is sufficient for our needs, as the primary constraint comes from the most 
massive halos in our catalogues.

We construct the distribution of linear overdensity $\delta$ as a function of the smoothing scale $\sigma(M)$
for the regions that collapse to form halos in the manner described in \S 
\ref{section:simulations:smoothed_overdensity},
using the same set of smoothed linear density fields calculated in \S 
\ref{section:simulations:smoothed_field}.  The mean of the $\delta(\sigma)$ 
distribution is calculated in three bins of width $\Delta\sigma=0.25$.  For the $\Delta=200$ halo
definition, the calculation results in the distribution of $\delta$ vs. $\sigma(M)$ shown in Figure 
\ref{fig:smoothed_overdensity} (at $\sigma(M)\lesssim1.2$).

To calculate an excursion set mass function from the overdensity distribution for each 
halo mass definition,
the mean overdensity $\delta$ as a function of $\sigma(M)$ is fit with two analytical
forms to produce two model collapse barriers.  We first fit the function
\begin{equation}
\label{equation:smt_barrier_fit}
\delta_{\mathrm{fit}} = \delta_{c}\left[1 + \beta(\sigma^{2}/\delta_{c}^{2})^{\gamma}\right],
\end{equation}
\noindent
used by \cite{sheth2001a} to represent ellipsoidal collapse, allowing $\beta$ and $\gamma$ to
vary.  By construction, this function converges to $\delta_{\mathrm{fit}}\to\delta_{c}$ as $\sigma\to0$.
We also fit a linear function,
\begin{equation}
\label{equation:linear_barrier_fit}
\delta_{\mathrm{lin}} = A\sigma + b,
\end{equation}
\noindent
allowing $A$ and $b$ to vary.  The intercept $b$ in general is smaller than $\delta_{c}$; 
this functional form thus allows us to test how results would change if the barrier does
not asymptote to $\delta_c$ for low values of $\sigma$.
The best fit parameters for the effective collapse barrier for
each halo mass definition are reported in Table \ref{table:barriers}.
We then use the best fit parameters for Equations
\ref{equation:smt_barrier_fit} and \ref{equation:linear_barrier_fit} to calculate a sharp $k$-space 
first-crossing distribution
via the method of \cite{zhang2006a} [i.e., Equation~(\ref{equation:zh06_volterra_fs})], and compare with
the simulated first-crossing distribution provided by the halo catalogue.

\begin{figure}
\figurenum{6a}
\epsscale{1.0}
\plotone{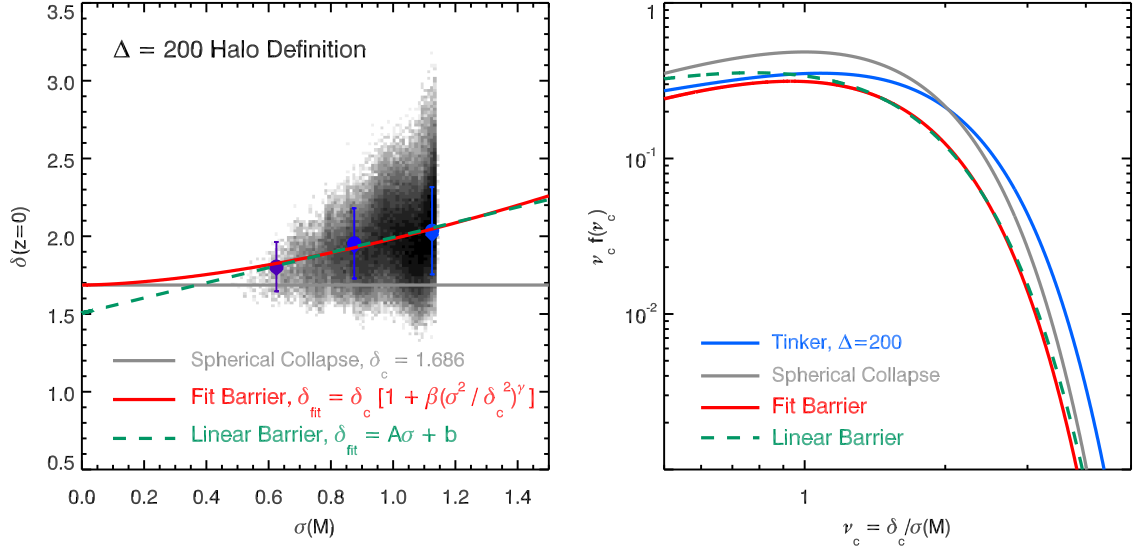}
\caption{\label{fig:dsf_200}
Smoothed linear overdensity $\delta$, extrapolated to $z=0$,  as a function of smoothing scale $\sigma(M)$ for regions
that collapse to form $\Delta=200$ halos by the present epoch (left panel).  
 The circles correspond to the mean overdensities, while the errorbars
 indicate the halo-to-halo scatter.  The error on the mean is
 significantly smaller than the scatter in all cases.  Shown for
 comparison is the spherical collapse barrier ($\deltac$, solid gray
 line).  Solid red
 line shows a fit of the functional form of the ellipsoidal collapse
barrier to the simulation results, while the dashed green line shows
a simple linear fit.  The right panel shows excursion set mass functions
 for each model barrier calculated using the method of
 \cite{zhang2006a} with sharp $k$ filter and compared with the spherical
 collapse (gray line, right panel) and \cite{tinker2008a} $\Delta=200$
 (blue line, right panel) mass functions.  \\\\\\ } \end{figure}

Figure \ref{fig:dsf_200} shows the distribution of linear overdensity $\delta$ with smoothing scale $\sigma(M)$
for the $\Delta=200$ halo catalogue.  As in Figure \ref{fig:smoothed_overdensity}, the mean overdensity of
regions that collapse to form halos in the $\Delta=200$ catalogue lie above the spherical collapse barrier $\deltac$.
Analytical fits to the mean of the overdensity distribution (i.e., the effective collapse barrier for this halo mass definition)
show that a simple linear extrapolation possibly suggests that $\delta\to1.5$ as $\sigma\to0$.  The excursion set first-crossing
distributions calculated for models of the mean overdensity in regions that form $\Delta=200$ halos show that the differences 
in the fits at $\sigma\lesssim0.45$ have little influence on the resulting halo abundance, as the larger $\delta>\deltac$ at
scales $\sigma\gtrsim 0.45$ suppresses the abundance of halos at $\nuc<4$ relative to the spherical collapse or \cite{tinker2008a}
$\Delta=200$ mass functions (see the right panel of Figure \ref{fig:dsf_200}).

\begin{figure}
\figurenum{6b}
\epsscale{1.0}
\plotone{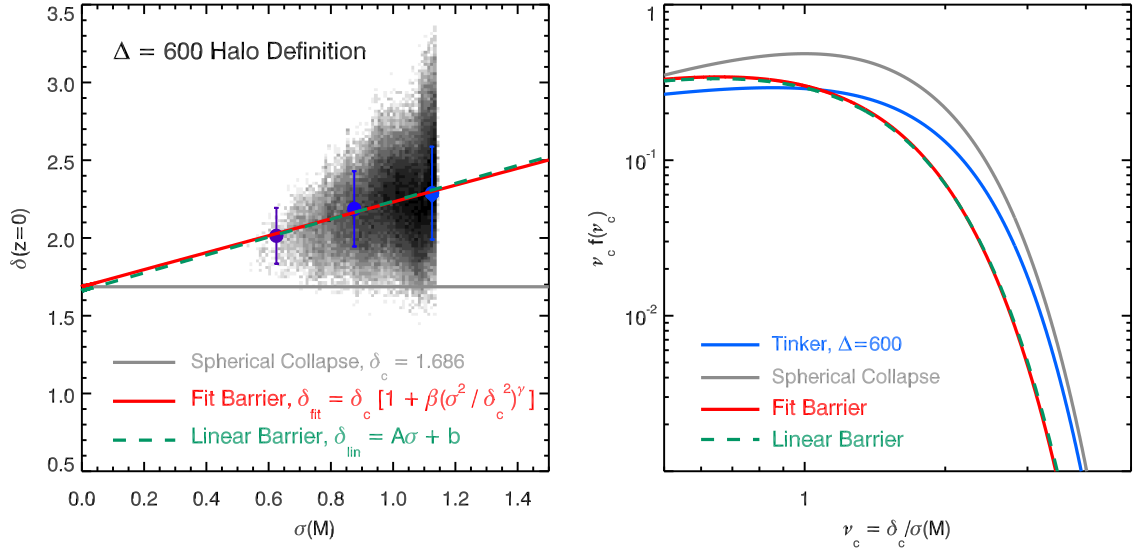}
\caption{\label{fig:dsf_600}
Same as for Figure \ref{fig:dsf_200}, but for halos defined with a $\Delta=600$ spherical overdensity criterion.
Increasing the overdensity threshold in the halo mass definition does not improve the agreement between the
excursion set and simulated halo mass function.
\\\\\\
}
\end{figure}

\begin{figure}
\figurenum{6c}
\epsscale{1.0}
\plotone{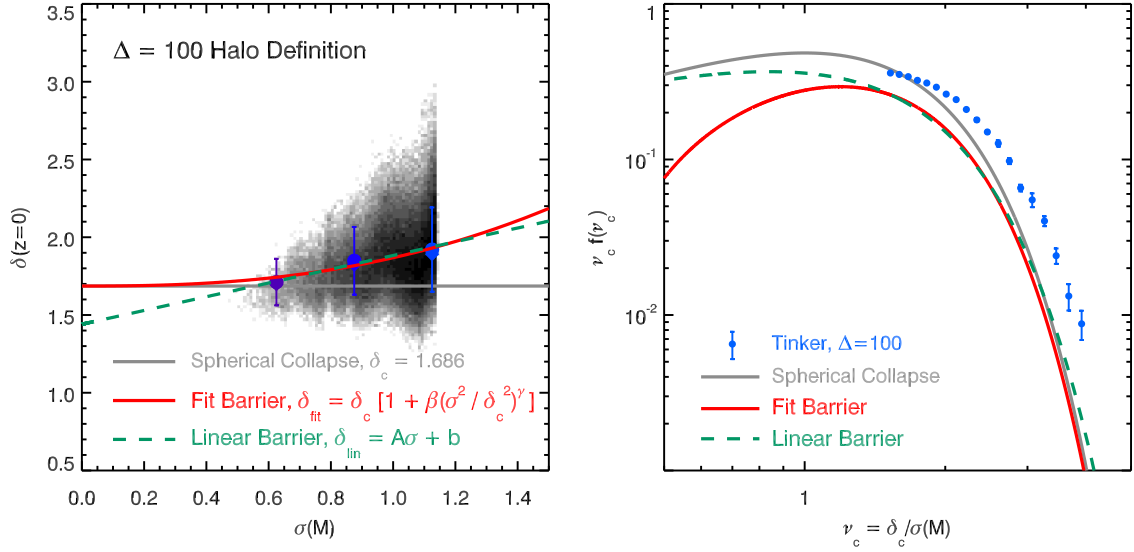}
\caption{\label{fig:dsf_100}
Same as for Figure \ref{fig:dsf_200}, but for halos defined with a $\Delta=100$ spherical overdensity criterion.
Here, the simulated halo mass function is measured directly from only the L1000W simulation (points and error bars).
\\\\\\
}
\end{figure}

Changing the overdensity threshold in the spherical overdensity halo definition has an intuitive effect on the linear overdensity of
regions that collapse to form halos.  A higher threshold overdensity, such as a $\Delta=600$ halo mass definition 
(Figure \ref{fig:dsf_600}), results in a higher characteristic linear overdensity for regions that collapse to form halos.  Model
fits to the mean overdensity with smoothing scale show that the mean overdensity increases roughly linearly, and approaches 
$\delta\to\deltac$ as $\sigma\to0$. The abundance of halos is correspondingly suppressed, with the model barrier fits producing very
similar excursion set first-crossing distributions that lie below the spherical collapse mass function.  The simulated $\Delta=600$
mass function from \cite{tinker2008a} has a lower abundance than the lower threshold $\Delta=200$ mass function, but the excursion
set mass functions are significantly lower than the simulated $\Delta=600$ mass function (Figure \ref{fig:dsf_600}, right panel).  
In this case, increasing the overdensity
threshold does not improve the performance of the excursion set {\it ansatz}.
Similarly, using a lower threshold such as a $\Delta=100$ halo mass definition (Figure \ref{fig:dsf_100}), does not bring the
excursion set and simulated mass functions into agreement.  The lower overdensity threshold decreases the characteristic linear
overdensity of regions that collapse to form halos and increases the abundance of regions that can collapse.  However, the mean
overdensity does not decrease significantly below the spherical collapse overdensity and the excursion set mass functions only
begin to roughly match abundance predicted by the spherical collapse mass function.  The simulated $\Delta=100$ mass function has increased
the abundance of halos relative to the $\Delta=200$ mass function, so the disagreement between the excursion set mass function and
the simulated mass function still remains for this lowered overdensity threshold (Figure \ref{fig:dsf_100}, right panel).

\begin{figure}
\figurenum{6d}
\epsscale{1.0}
\plotone{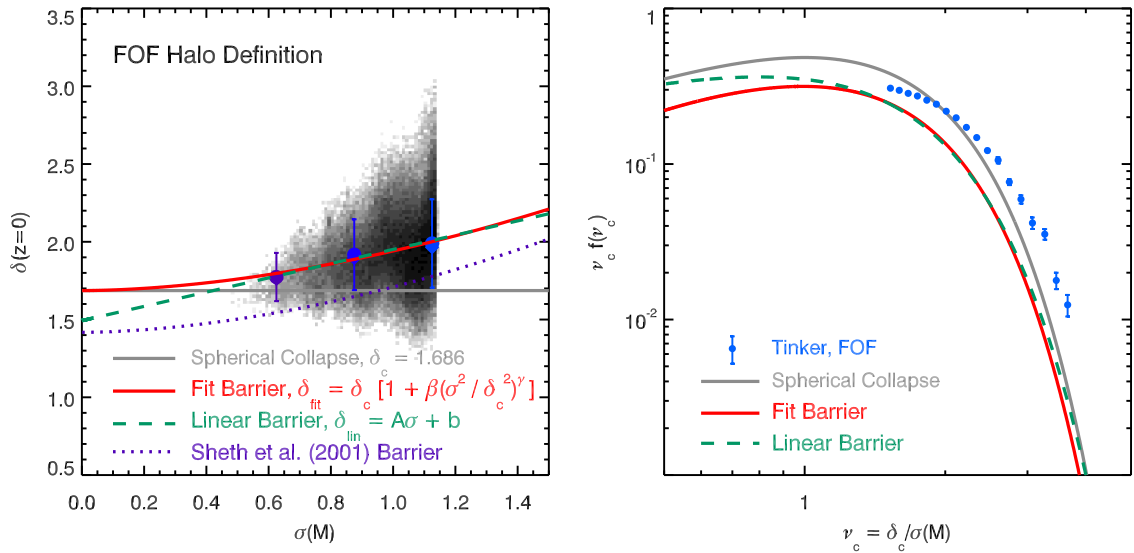}
\caption{\label{fig:dsf_fof}
Same as for Figure \ref{fig:dsf_100}, but for halos defined with a Friends-of-Friends (FOF) algorithm using a 
linking length of $b=0.2$. 
The characteristic linear overdensity of regions that collapse to form FOF halos does not follow
the modified ellipsoidal collapse barrier presented by \cite[][purple dotted line]{sheth2001a}. 
Changing the halo definition from spherical overdensity to FOF halos does not improve the agreement between the
excursion set and simulated halo mass function.
The simulated halo mass function is measured directly from only the L1000W simulation (points and error bars).
\\\\\\
}
\end{figure}

Altering the mass definition from spherical overdensity to Friends-of-Friends (FOF) halos does not improve the agreement
between the simulated and excursion set mass function.  The original motivation presented by \cite{sheth2001a} for 
modifying the ellipsoidal collapse barrier to the lower limiting value of $\delta_{\mathrm{SMT}}\to\sqrt{\aSMT}\deltac$ as
$\sigma\to0$ was the use of a FOF definition in identifying halos in the GIF simulation.  
Figure \ref{fig:dsf_fof} demonstrates explicitly that the
characteristic linear overdensity of regions that collapse to form FOF halos does not follow Equation 
\ref{equation:gif_simulation_barrier}; the mean overdensity is similar to that found for $\Delta=200$ halos.
Similarly, the excursion set mass functions calculated from the model fits to the mean overdensity of regions
that collapse to form FOF halos do not match the simulated FOF halo mass function.

\begin{figure}
\figurenum{6e}
\epsscale{1.0}
\plotone{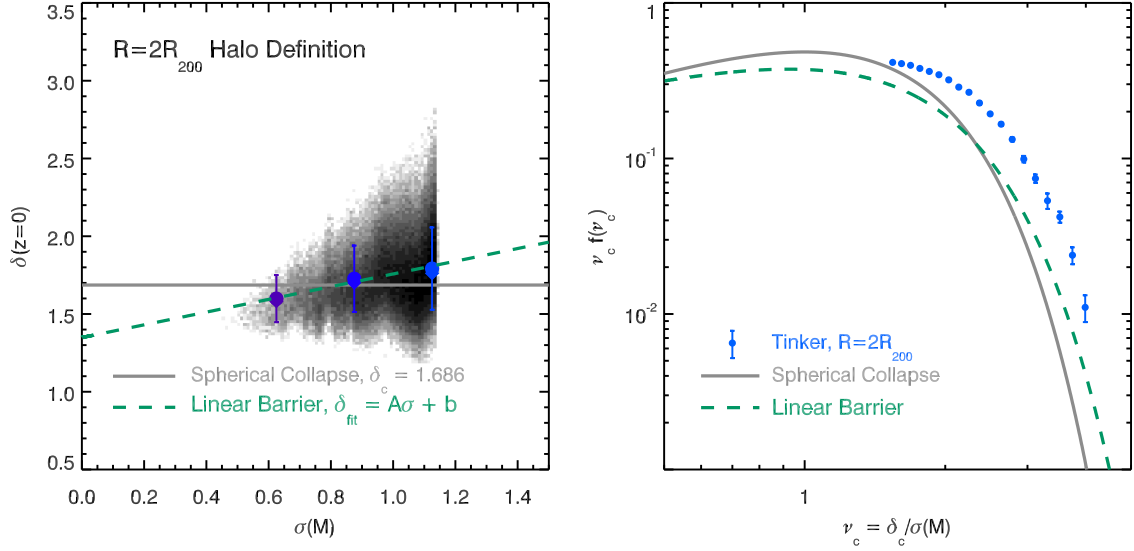}
\caption{\label{fig:dsf_2Rvir}
Same as for Figure \ref{fig:dsf_100}, but for $\Delta=200$ halos with masses rescaled to include all particles
within a radius $R=2R_{200}$.  In this halo definition, small $\Delta=200$ halos with centers-of-mass
that reside within a distance $R$ of larger halos are incorporated into the larger system.
The mean of the linear overdensity distribution 
in this case lies below the spherical collapse barrier at small $\sigma$, so only the excursion set mass function 
for the linear barrier (which allows $\delta<\deltac$) is compared with the simulated mass 
function (right panel).
Changing the halo mass definition to increase the region incorporated into halos identified by a $\Delta=200$
spherical overdensity criterion does not improve the agreement between the
excursion set and simulated halo mass function (points and error bars).
\\\\\\
}
\end{figure}

We could also redefine the mass from the $\Delta=200$ halo definition to include all particles within a modified ``virial'' radius 
$R=2R_{200}$.  This halo mass definition is intended as an analogy to the ``static'' mass halo definition proposed by
\cite{cuesta2008a}, who found that halos defined by regions with zero mean radial velocity 
(with a size of approximated $2\times R_{\vir}$, see their Figure 14) displayed an abundance similar to the Press-Schechter
spherical collapse mass function at $z=0$.  
This halo definition results in halo abundance roughly twice that
found by \cite{cuesta2008a} at fixed mass because the effective $R_{\vir}/R_{200}$ ratio is mass-dependent, but
provides a useful example of a mass definition that incorporates very large regions into single halos.
Figure \ref{fig:dsf_2Rvir} shows that the linear overdensity of regions that
collapse to form halos defined in this manner is typically low, since the typical overdensity at $2R_{200}$ 
is quite low, and decreases below the spherical collapse barrier at
small $\sigma$.  Hence, we only fit the mean overdensity with a linear barrier model that allows for $\delta<\deltac$
and do not report the best fit parameters for the model defined by Equation \ref{equation:smt_barrier_fit}.
As is clear from the right panel of Figure \ref{fig:dsf_2Rvir}, the excursion set mass function 
calculated for the linear barrier model fit for the $R=2R_{200}$
halos does not recover the simulated halo mass function and, therefore, this halo mass definition does not
improve the success of the excursion set {\it ansatz}.

\begin{figure}
\figurenum{7}
\epsscale{0.5}
\plotone{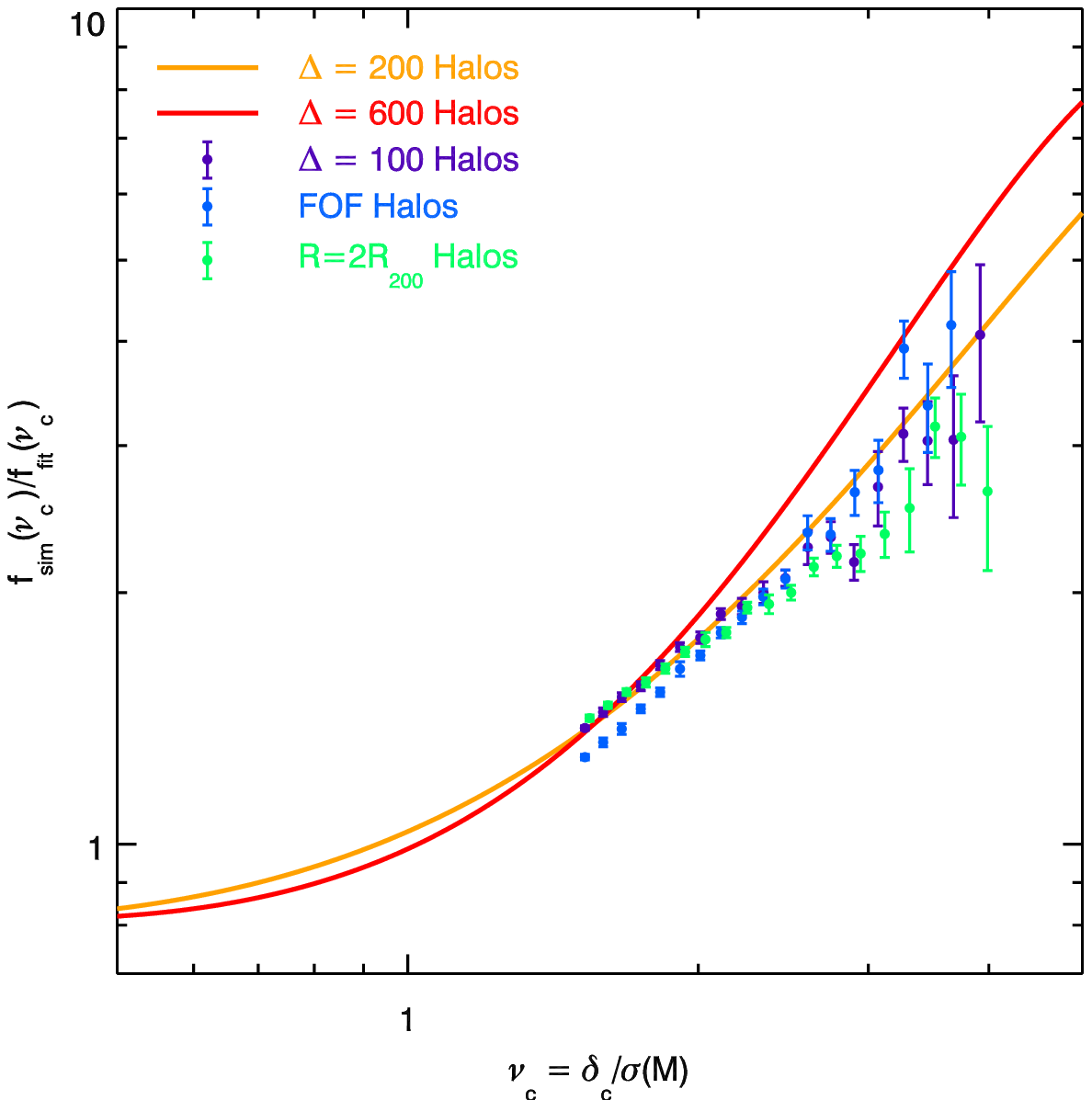}
\caption{\label{fig:mass_def_normalization}
Ratio $f_{\mathrm{sim}}(\nuc)/f_{\mathrm{fi}}(\nuc)$ of the simulated first-crossing distribution $f_{\mathrm{sim}}$
to the excursion set first-crossing distribution $f_{\mathrm{fit}}$ calculated from a linear fit to the overdensity 
distribution plotted in Figures \ref{fig:dsf_200}-\ref{fig:dsf_2Rvir} for a variety of halo mass definitions, plotted
as a function of peak height $\nuc=\deltac/\sigma(M)$.  
For the $\Delta=200$ (orange line) and $\Delta=600$ (red line) spherical overdensity halo definitions, the analytical
fits from \cite{tinker2008a} are compared with the calculated first-crossing distributions.  For $\Delta=100$
spherical overdensity halos (purple points), Friends-of-Friends halos (blue points), 
and halos with masses defined by the particle content within
a radius $R=2R_{200}$ (green points), the binned first-crossing distributions from the L1000W simulation halo 
catalogue were compared with the excursion set results at the appropriate $\nuc$ for each mass bin.  For these
halo mass definitions, the
uncertainty estimates in $f_{\mathrm{sim}}/f_{\mathrm{fit}}$ reflect the same fractional uncertainty in the
simulated first-crossing distributions shown in Figures \ref{fig:dsf_100}-\ref{fig:dsf_2Rvir}.  
For every halo
mass definition, that $f_{\mathrm{sim}}/f_{\mathrm{fit}}$ is not constant with $\nuc$ demonstrates that the
disagreement between the simulated and excursion set first-crossing distributions differ in shape and
not simply in relative normalization.
\\\\\\
}
\end{figure}

Lastly, as in the discussion in \S \ref{section:simulations}, one might wonder whether the disagreement between the 
simulation and excursion set mass functions simply involves a normalization issue.  In fact, the normalization and
shape of each of the excursion set mass functions differ from the mass function constructed from the simulated halo
abundance.  Figure \ref{fig:mass_def_normalization} shows $f_{\mathrm{sim}}(\nuc)/f_{\mathrm{fit}}(\nuc)$, the ratio 
of the first-crossing distribution $f_{\mathrm{sim}}(\nuc)$ determined from the simulations to $f_{\mathrm{fit}}(\nuc)$,
the first-crossing distribution calculated using the linear barrier model fits to the overdensity distribution, for
each of the halo mass definitions considered in this paper ($\Delta=100$, $\Delta=200$, and $\Delta=600$ spherical
overdensities, FOF halos, and halos with masses determined by the particle distribution within a radius $R=2R_{200}$ of
the halo center-of-mass).  For the $\Delta=200$ and $\Delta=600$ halo definitions, Figure \ref{fig:mass_def_normalization}
shows the ratio of the best fit analytical \cite{tinker2008a} and excursion set first-crossing distributions.  For the other
halo definitions, the binned first-crossing distribution determined by the simulated halo mass function is divided
by the excursion set first-crossing distribution at the appropriate $\nuc$ value.  Uncertainty estimates for 
$f_{\mathrm{sim}}(\nuc)/f_{\mathrm{fit}}(\nuc)$ for these halo mass definitions (the error bars in Figure 
\ref{fig:mass_def_normalization}) have the same fractional error as for the first-crossing distributions plotted
in Figures \ref{fig:dsf_100}-\ref{fig:dsf_2Rvir}.
For each mass definition, $f_{\mathrm{sim}}/f_{\mathrm{fit}}$ varies with the peak height $\nuc$ and 
demonstrates that the
disagreement between the simulated and excursion set first-crossing distributions 
does not owe simply to their relative normalization.

\begin{deluxetable*}{|l|cc|cc|} 
\tablecolumns{5} 
\tablewidth{0pc} 
\tablecaption{Best Fit Barrier Model Parameters} 
\tablehead{
\colhead{Mass Definition} & \colhead{$\beta$} & \colhead{$\gamma$} & \colhead{$A$} & \colhead{$b$}
\label{table:barriers}}
\startdata 
$\Delta=100$ & 0.396 & 1.242 & 0.442  & 1.441 \\
$\Delta=200$ & 0.411 & 0.809 & 0.487  & 1.506 \\
$\Delta=600$ & 0.543 & 0.496 & 0.576  & 1.660 \\
FOF          & 0.363 & 0.890 & 0.458  & 1.493 \\
$R_{\vir}=2R_{\Delta=200}$ & \nodata & \nodata & 0.409  & 1.348 \\
\enddata 
\end{deluxetable*}

\end{document}